\documentclass[11pt,a4paper,twoside]{article}
\usepackage{graphicx}
\usepackage{color}
\usepackage{hyperref}
\usepackage{setspace}
\usepackage{epstopdf}
\begin{document}
\baselineskip=0.20in
\vspace{20mm}
\baselineskip=0.30in
{\bf \LARGE
\begin{center}
Relativistic symmetries in trigonometric P\"oschl-Teller potential plus tensor interaction
\end{center}}
\vspace{6mm}
\begin{center}
{\Large {\bf Babatunde J. Falaye}}\footnote{\scriptsize E-mail:~ fbjames11@physicist.net} \\
\small
\vspace{2mm}
{\it Theoretical Physics Section, Department of Physics\\
 University of Ilorin,  P. M. B. 1515, Ilorin, Nigeria.}
\end{center}
\vspace{2mm}
\begin{center}
{\Large {\bf Sameer M. Ikhdair}}\footnote{\scriptsize E-mail:~ sikhdair@neu.edu.tr} \\
\small
\vspace{2mm}
{\it Department of Physics, Faculty of Science, An-Najah National University, \\
New campus, Junaid, Nablus, West Bank, Palestine,}\\
and\\
{\it Department of Electrical and Electronic Engineering, Near East University, \\ 922022 Nicosia, Northern Cyprus, Mersin 10, Turkey.}
\end{center}
\vspace{12mm}
\noindent
\begin{abstract}
\noindent
The Dirac equation is solved to obtain its approximate bound states for a spin-$1/2$ particle in the presence of trigonometric P\"oschl-Teller (tPT) potential including a Coulomb-like tensor interaction with arbitrary spin-orbit quantum number $\kappa$  using an approximation scheme to substitute the centrifugal terms $\kappa(\kappa\pm 1)r^{-2}$. In view of spin and pseudo-spin (p-spin) symmetries, the relativistic energy eigenvalues and the corresponding two-component wave functions of a particle moving in the field of attractive and repulsive tPT potentials are obtained using the asymptotic iteration method (AIM). We present numerical results in the absence and presence of tensor coupling $A$ and for various values of spin and p-spin constants and quantum numbers $n$ and $\kappa$. The non-relativistic limit is also obtained.
\end{abstract}


\section{Introduction}
The concept of pseudo-spin (p-spin) and spin symmetries in nucleon and antinucleon, respectively, introduced to reveal the dynamical nature of quantum systems and these symmetries play a critical role in the shell structure and its evolution. Within the framework of Dirac equation, the p-spin symmetry is used to feature the deformed nuclei and the superdeformation to establish an effective shell-model \cite{BJ1, BJ2,BJ3, BJ4}, whereas spin symmetry is relevant for mesons \cite{BJ5}. The spin symmetry can be regarded as a relativistic symmetry. For the p-spin, in nuclei, its origin has not been fully clarified until now \cite{BJ6, BJ7, BJ8}. The exact spin symmetry occurs when the scalar $S(r)$ and vector $V(r)$ potentials have nearly the same size but same sign, i.e., $\Delta(r)=S(r)-V(r)=C_s=0$ and the exact p-spin symmetry occurs when the scalar and vector potentials have the same size but opposite sign, i.e., $\Sigma(r)=S(r)+V(r)=C_{ps}=0$ \cite{BJ6, BJ7,BJ8}. 

Later it was found that spin and p-spin are exact $\frac{d}{dr}\Delta(r)$ and $\frac{d}{dr}\Sigma(r)$, respectively, under a less strict condition and to what extent the spin and p-spin are conserved are related to the competition between the centrifugal barrier and the spin or p-spin orbital potential \cite{BJ6, BJ7,BJ8}. The condition can not be met in realistic nuclei. For more details, see the recent works \cite{BJ6, BJ7,BJ8, BJ9, BJ10} and references therein. The p-spin symmetry refers to a quasi-degeneracy of single nucleon doublets with non-relativistic quantum number $(n, \ell, j=\ell+1/2)$ and $(n-1,\ell+2,j=\ell+3/2)$, where $n$, $\ell$ and $j$ are single nucleon radial, orbital and total angular quantum numbers, respectively \cite{BJ11, BJ12,BJ13, BJ14}. The total angular momentum is $j=\tilde{\ell}+\tilde{s}$, where $\tilde{\ell}=\ell+1$  pseudo-angular momentum and $\tilde{s}$ is p-spin angular momentum \cite{BJ15}. Very recently, the tensor potential was introduced into the Dirac equation by simply substituting $\vec{P}\rightarrow \vec{P}-im\omega\beta.\hat{r}U(r)$ and a spin-orbit coupling is added to the Dirac Hamiltonian $[16-26]$. 

The unbound solutions of generalized asymmetrical Hartmann potentials under the condition of the p-spin symmetry have been presented by mapping the wave functions of bound states in the complex momentum plane via the continuation method \cite{BJ9}.
Very recently, Guo explored the p-spin symmetry by using the similarity renormalization group and shown explicitly the relativistic origin of the symmetry \cite{BJ7}. Chen and Guo \cite{BJ8} investigate the evolution of the spin and pseudospin symmetries
from the relativistic to the nonrelativistic and explore the relativistic relevance of the symmetries. By examining the zeros of Jost functions corresponding to the small components of Dirac wave functions and phase shifts of continuum states, Lu et al \cite{BJ6} show that the pseudospin symmetry in single particle resonant states in nuclei is conserved when the attractive scalar and repulsive vector potentials have the same magnitude but opposite sign.

In the present work, we intend to investigate the trigonometric P\"oschl-Teller (tPT) potential proposed for the first time by P\"oschl and Teller \cite{BJ27} in 1933 was to describe the diatomic molecular vibration. Chen \cite{BJ28} and Zhang et al. \cite{BJ29} have studied the relativistic bound state solutions for the tPT potential and hyperbolical PT (Second PT) potential, respectively. Liu et al. \cite{BJ30} studied the tPT potential within the framework of the Dirac theory. Recently, Candemir investigated the analytic $\tilde{s}-$wave solution of Dirac equation for tPT potential under the p-spin symmetry condition \cite{BJ31}. Further, Hamzavi and Rajabi studied the exact $s-$wave  $(\ell=0)$ solution of the Schr\"odinger equation for the vibration tPT potential \cite{BJ32}. Very recently, Hamzavi et al. investigated the approximate analytic bound state eigensolutions of the Dirac equation with the tPT potential including the centrifugal (pseudo-centrifugal) term for any spin-orbit quantum number $\kappa$ in view of spin and p-spin symmetries \cite{BJ33}. The energy spectrum of the tPT via the asymptotic iteration method have been obtained recently by Falaye \cite{BJ37}. This potential takes the following form:
\begin{equation}
V(r)=\frac{V_1}{\sin^2(\alpha r)}+\frac{V_2}{\cos^2(\alpha r)},
\label{POT}
\end{equation}
where the parameters $V_1$  and $V_2$ describe the property of the potential well while the parameter $\alpha$ is related to the range of this potential \cite{BJ30}. In Figures $1$ and $2$, we draw the tPT potential (\ref{POT}) for parameter values $V_1=5.0fm^{-1}$, $V_2=3.0fm^{-1}$, $\alpha=0.02fm^{-1}$ and $\alpha=0.03fm^{-1}$. 

The goal of the present work is to extend previous works \cite{BJ31,BJ32,BJ33,BJ37} to the relativistic case, $\kappa\neq\pm 1$ (rotational case) and also in the presence of a Coulomb-like tensor potential via the asymptotic iteration method \cite{BJ34,BJ35,BJ36,BJ37}. We introduce a convenient approximation scheme to deal with the strong singular centrifugal term. The ansatze of this approximation possesses the same form of the potential and is singular as the centrifugal term $r^{-2}$  \cite{BJ38}. 
   
The structure of the paper is as follows. In Section \ref{sec2}, we briefly introduce AIM method. In Section \ref{sec3}, in view of spin and p-spin symmetries, we briefly introduce the Dirac equation with scalar and vector tPT potentials plus a Coulomb-like tensor interaction for arbitrary spin-orbit quantum number. In Section \ref{sec4}, the approximate energy eigenvalue equations and corresponding two-component wave functions of the Dirac-tPT problem including tensor interaction are obtained. The non-relativistic limit of our solution is also obtained. We present our numerical results in the presence and absence of tensor coupling $A$ for various $n$ and $\kappa$ quantum numbers. We end with our concluding remarks in Section \ref{sec5}.
\section{Method of Analysis}
\label{sec2}
One of the calculational tools utilized in solving the Schr\"{o}dinger-like equation including the centrifugal barrier and/or the spin-orbit coupling term is called as the asymptotic iteration method (AIM). For a given potential the idea is to convert the Schr$\ddot{o}$dinger-like equation to the homogenous linear second-order differential equation of the form \cite{BJ34}:
\begin{equation}
y''(x)=\lambda_o(x)y'(x)+s_o(x)y(x),
\label{E1}
\end{equation}
where $\lambda_o(x)$ and $s_o(x)$ have sufficiently many continous derivatives and defined in some interval which are not necessarily bounded. The differential equation (\ref{E1}) has a general solution $\cite{BJ34,BJ35}$
\begin{equation}
y(x)=\exp\left(-\int^x\alpha(x')dx'\right)\left[C_2+C_1\int^x\exp\left(\int^{x'}\left[\lambda_o(x'')+2\alpha(x'')\right]dx''\right)dx'\right].
\label{E2}
\end{equation}
If $k>0$, for sufficiently large $k$, we obtain the $\alpha(x)$
\begin{equation}
\frac{s_k(x)}{\lambda_k(x)}=\frac{s_{k-1}(x)}{\lambda_{k-1}(x)}=\alpha(x) \ ,\ \ k=1, 2, 3.....
\label{E3}
\end{equation}
where
\begin{eqnarray}
\lambda_k(x)&=&\lambda'_{k-1}(x)+s_{k-1}(x)+\lambda_o(x)\lambda_{k-1}(x),\nonumber\\
s_k(x)&=&s'_{k-1}(x)+s_o(x)\lambda_{k-1}(x) \ ,\ \ k=1, 2, 3.....
\label{E4}
\end{eqnarray}
The energy eigenvalues are obtained from the quantization condition of the method together with equation (\ref{E4}) and can be written as follows:
\begin{equation}
\delta_k(x)=\lambda_k(x)s_{k-1}(x)-\lambda_{k-1}(x)s_{k}(x)=0\ \ ,\ \  \ k=1, 2, 3....
\label{E5}
  \end{equation}
The energy eigenvalues are then obtained from (\ref{E5}), if the problem is exactly solvable.  If not, for a specific $n$ principal quantum number, we choose a suitable $x_0$ point, determined generally as the maximum value of the asymptotic wave function or the minimum value of the potential and the aproximate energy eigenvalues are obtained from the roots of this equation for sufficiently large values of $k$ with iteration. 
\section{Dirac Equation with Scalar and Vector Potentials Including Tensor Coupling Interaction}
\label{sec3}
In spherical coordinates, the Dirac equation for fermonic massive spin$-\frac{1}{2}$ particles interacting with arbitrary scalar potential S(r), the time-component $V(r)$ of a four-vector potential and the tensor potential $U(r)$ can be expressed as \cite{BJ5, BJ6} 
\begin{equation}
\left[\vec{\alpha}.\vec{p}+\beta(M+S(r))-i\beta\vec{\alpha}.\hat{r}U(r)\right]\psi(\vec{r})=[E-V(r)]\psi(\vec{r}),
\label{E6}
\end{equation}
where $E$, $\vec{p}$ and $M$ denote the relativistic energy of the system, the momentum operator and mass of the particle respectively. $\alpha$ and $\beta$ are $4\times 4$ Dirac matrices given by
\begin{equation}
\bar{\alpha}=
\left(\begin{array}{lr}     
0&\vec{\sigma }\\      
\vec{\sigma }&0
\end{array}\right)\ ,\ \ \ \beta=
\left(\begin{array}{lr}     
I&0\\      
0&-I
\end{array}\right),\ \ \ {\sigma_1}=
\left(\begin{array}{lr}     
0&1\\      
1&0
\end{array}\right),\ \ \ {\sigma_2}=
\left(\begin{array}{lr}     
0&-1\\      
i&0
\end{array}\right),\ \ \ {\sigma_3}=
\left(\begin{array}{lr}     
1&0\\      
0&-1
\end{array}\right),
\label{E7}
\end{equation}
where $I$ is the $2\times2$ unitary matrix and $\vec{\sigma}$  are the three-vector pauli spin matrices. The eigenvalues of the spin-orbit coupling operator are $\kappa=\left(j+\frac{1}{2}\right)>0$ and $\kappa=-\left(j+\frac{1}{2}\right)<0$ for unaligned spin $j=\ell-\frac{1}{2}$ and the aligned spin $j=\ell+\frac{1}{2}$ respectively. The set $(H^2, K, J^2, J_Z)$ can be taken as the complete set of conservative quantities with $\vec{J}$ being the total angular momentum operator and $K=(\vec{\sigma} .\vec{L}+1)$ is the spin-orbit where $\vec{L}$ is the orbital angular momentum of the spherical nucleons that commutes with the Dirac Hamiltonian. Thus, the spinor wave functions can be classified according to their angular momentum $j$, the spin-orbit quantum number $\kappa$ and the radial quantum number $n$. Hence, they can be written as follows:
\begin{equation}
\psi_{n\kappa}(\vec{r})=
\left(\begin{array}{lr}     
f_{n\kappa}(\vec{r})\\      
g_{n\kappa}(\vec{r})
\end{array}\right)=\frac{1}{r}\left(\begin{array}{lr}     
F_{n\kappa}(\vec{r})Y^\ell_{jm}(0,\phi)\\      
iG_{n\kappa}(\vec{r})Y^{\tilde{\ell}}_{jm}(0,\phi),
\end{array}\right),
\label{E8}
\end{equation}
where $F_{n\kappa}(\vec{r})$ and $G_{n\kappa}(\vec{r})$ are the radial wave functions of the upper- and lower-spinor components respectively and $Y^\ell_{jm}(0,\phi)$ and $Y^{\tilde{\ell}}_{jm}(0,\phi)$ are the spherical harmonic functions coupled to the total angular momentum $j$ and it's projection $m$ on the $z$ axis. Substitution of equation (\ref{E6}) into equation (\ref{E1}) yields the following coupled differential equations \cite{N1, N2, N3, N4, N5, N6, N7}:
\begin{eqnarray}
\left(\frac{d}{dr}+\frac{\kappa}{r}-U(r)\right)F_{n\kappa}(r)&=&(M-E_{n\kappa}-\Delta(r))G_{n\kappa}(r)\nonumber\\
\left(\frac{d}{dr}-\frac{\kappa}{r}+U(r)\right)G_{n\kappa}(r)&=&(M-E_{n\kappa}-\Sigma(r))F_{n\kappa}(r)
\label{E9}
\end{eqnarray}
where $\Delta(r)=V(r)-S(r)$ and $\Sigma(r)=V(r)+S(r)$ are the difference and sum potentials respectively. On solving in equation (\ref{E9}), we obtain the following Schr$\ddot{o}$diger-like differential equation with coupling to the $r^{-2}$ singular term and satisfying $F_{n\kappa}(r)$:
\begin{eqnarray}
\left[\frac{d^2}{dr^2}-\frac{\kappa(\kappa+1)}{r^2}+\frac{2\kappa}{r}U(r)-U^2(r)-\frac{dU(r)}{dr}+\frac{\frac{d\Delta(r)}{dr}}{M+E_{n\kappa}-\Delta(r)}\left(\frac{d}{dr}+\frac{\kappa}{r}-U(r)\right)\right]F_{n\kappa}(r)\nonumber\\
=\left[\left(M+E_{n\kappa}-\Delta(r)\right)\left(M-E_{n\kappa}+\Sigma(r)\right)\right]F_{n\kappa}(r),
\label{E10}
\end{eqnarray}
where $E_{n\kappa}\neq-M$ when $\Delta(r)=0$ and $\kappa(\kappa+1)=\ell(\ell+1)$. Since $E_{n\kappa}=+M$ is an element of the positive energy spectrum of the Dirac Hamiltonian, this relation with upper spinor component is not valid for the negative energy spectum solution. Furthermore, a similar equation satisfying $G_{n\kappa}(r)$ can be obtained as:
\begin{eqnarray}
\left[\frac{d^2}{dr^2}-\frac{\kappa(\kappa-1)}{r^2}+\frac{2\kappa}{r}U(r)-U^2(r)+\frac{dU(r)}{dr}+\frac{\frac{d\Sigma(r)}{dr}}{M-E_{n\kappa}+\Sigma(r)}\left(\frac{d}{dr}-\frac{\kappa}{r}+U(r)\right)\right]G_{n\kappa}(r)\nonumber\\
=\left[\left(M+E_{n\kappa}-\Delta(r)\right)\left(M-E_{n\kappa}+\Sigma(r)\right)\right]G_{n\kappa}(r),
\label{E11}
\end{eqnarray}
where $E_{n\kappa}\neq+M$ when $\Sigma(r)=0$ and $\kappa(\kappa-1)=\tilde{\ell}(\tilde{\ell}+1)$. Since $E_{n\kappa}=-M$ is an element of the negative energy spectrum of the Dirac Hamiltonian, this relation with the lower spinor component is not valid for the positive energy spectrum solution. 
\subsection{P-spin symmetry limit}
\label{PSL}
The p-spin symmetry occurs when $\frac{d[V(r)+S(r)]}{dr}=\frac{d\Sigma(r)}{dr}=0$ or $\Sigma(r)=C_{ps}=$constant. Here we are taking $\Delta(r)$ as the tPT potential and the tensor as the Coulomb-like potential, i.e.
\begin{equation}
\Delta(r)=\frac{V_1}{\sin^2\alpha r}+\frac{V_2}{\cos^2\alpha r}\ \ and \ \ U(r)=-\frac{A}{r}, \ \ \ \ r\geq R_c,
\label{E12}
\end{equation}
with
\begin{equation}
A=\frac{Z_aZ_be^2}{4\pi\epsilon_o},
\label{E13}
\end{equation}

\begin{figure}[!t]
\centering \includegraphics[height=100mm,width=140mm]{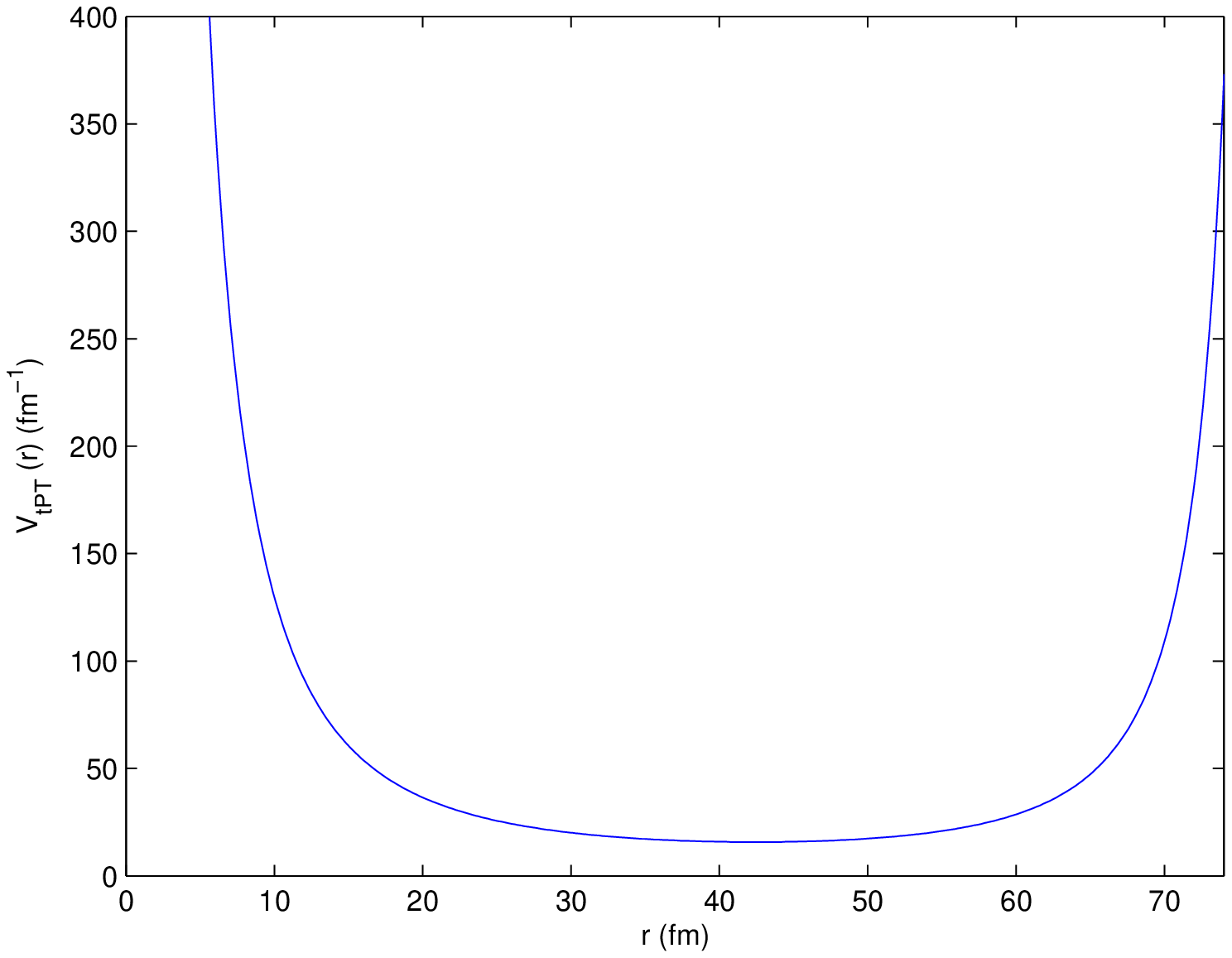}
\caption{{\protect\footnotesize The plot of the tPT as a function of $r$ for $\alpha=0.02fm^{-1}$}}
\label{fig1}
\end{figure}

\begin{figure}[!t]
\centering \includegraphics[height=100mm,width=140mm]{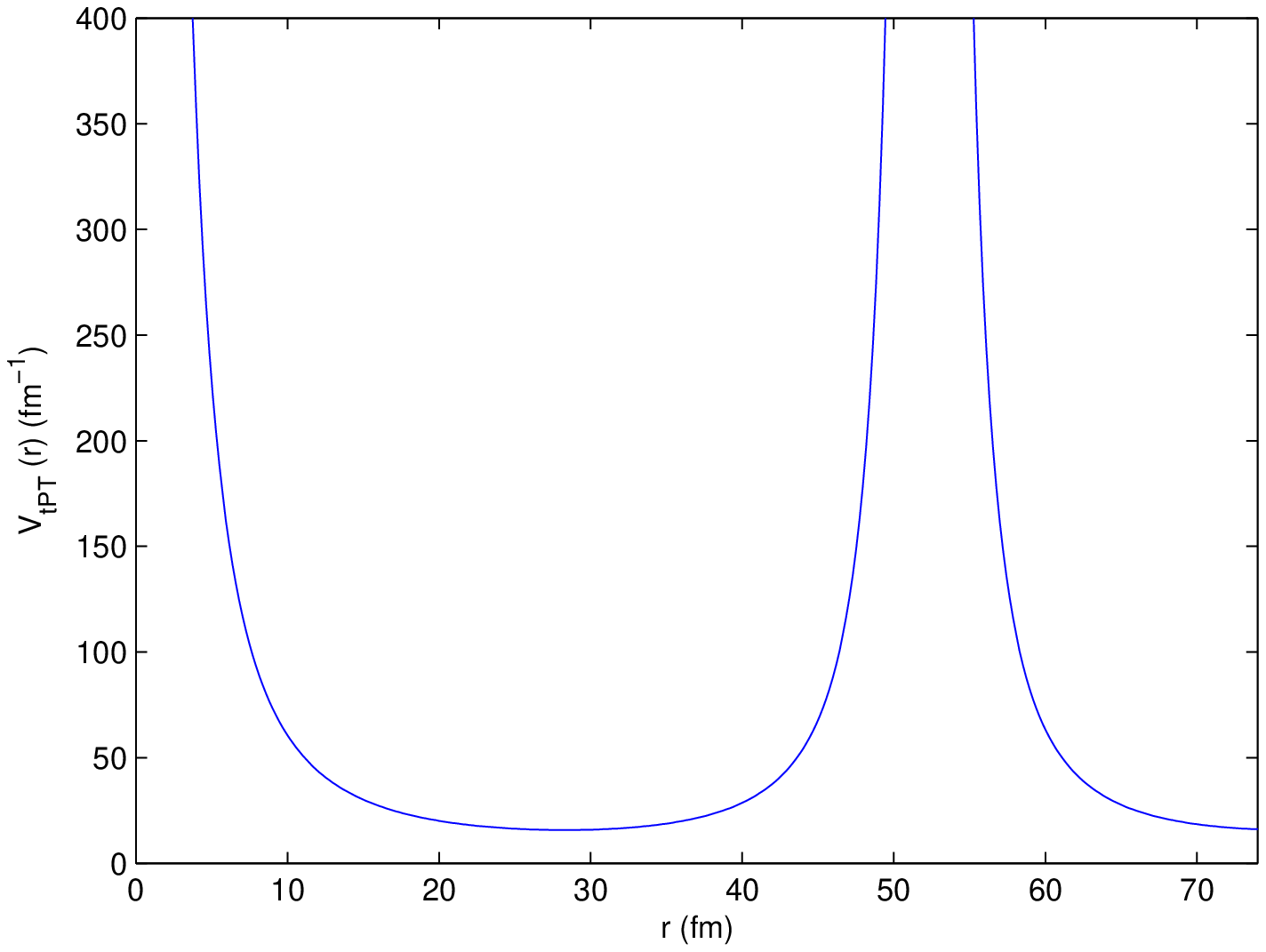}
\caption{{\protect\footnotesize The plot of the tPT as a function of $r$ for $\alpha=0.03fm^{-1}$}}
\label{fig2}
\end{figure}

where $R_c$ is the Coulomb radius, $Z_a$ and $Z_b$ respectively, denote the charges of the projectile $a$ and the target nuclei $b$ $\cite{BJ2, BJ3}$. Under this symmetry, equation (\ref{E11}) can easily be transformed to
\begin{equation}
\left[\frac{d^2}{dr^2}-\frac{\bar{\delta}}{r^2}-\frac{\bar{\gamma_1}}{\sin^2\alpha r}-\frac{\bar{\gamma}_2}{\cos^2\alpha r}-\bar{\beta}^2\right]G_{n\kappa}(r)=0,
\label{E14}
\end{equation}
where $\kappa=-\tilde{\ell}$ and $\kappa=\tilde{\ell}+1$ for $\kappa<0$ and $\kappa>0$ respectively and
\begin{eqnarray}
\bar{\gamma_1}&=&(E_{n\kappa}-M-C_{ps})V_1,\nonumber\\
\bar{\gamma_2}&=&(E_{n\kappa}-M-C_{ps})V_2,\nonumber\\
\bar{\beta}&=&\sqrt{(M+E_{n\kappa})(M-E_{n\kappa}+C_{ps})},\\ \bar{\delta}&=&(\kappa+A)(\kappa+A-1),\nonumber
\label{E15}
\end{eqnarray}
have been introduced for mathematical simplicity.
\subsection{Spin symmetry limit}
\label{SSL}
In the spin symmetry limit, $\frac{d\Delta(r)}{dr}=0$ or $\Delta(r)=C_s=$ constant $\cite{BJ5, BJ6}$. Similarly to section \ref{PSL}, we consider
\begin{equation}
\Sigma(r)=\frac{V_1}{\sin^2\alpha r}+\frac{V_2}{\cos^2\alpha r}\ \ and\ \  U(r)=-\frac{A}{r}, \ \ \ \ r\geq R_c.
\label{E16}
\end{equation}
Using equation (\ref{E16}), we can rewrite equation (\ref{E10}) as
\begin{equation}
\left[\frac{d^2}{dr^2}-\frac{{\delta}}{r^2}-\frac{\gamma_1}{\sin^2\alpha r}-\frac{\gamma_2}{\cos^2\alpha r}-{\beta}^2\right]F_{n\kappa}(r)=0,
\label{E17}
\end{equation}
where $\kappa=\ell$ and $\kappa=-\ell-1$ for $\kappa<0$ and $\kappa>0$, respectively. We have also introduced the following parameters
\begin{eqnarray}
\gamma_1&=&(M+E_{n\kappa}-C_{s})V_1,\nonumber\\ 
\gamma_2&=&(M+E_{n\kappa}-C_{s})V_2,\nonumber\\ 
\beta&=&\sqrt{(M-E_{n\kappa})(M+E_{n\kappa}-C_{s})},\\
\delta&=&(\kappa+A)(\kappa+A+1),\nonumber
\label{E18}
\end{eqnarray}
for mathematical simplicity.
\section{Approximate Relativistic Bound-States}
\label{sec4}
In this section, we present the approximate bound state solutions of the Dirac equation with the tPT potential in the presence of the tensor interaction using the AIM.
\subsection{Approximate bound-states for the p-spin symmetric limit}
It is clearly seen that equation (\ref{E14}) is not analytically solvable because of the coupling term. We therefore must include an approximation to deal with the coupling term. It is found that for short range potential, the following formula is a good approximation to the coupling term \cite{BJ30, BJ31, BJ32}:
\begin{equation}
\frac{1}{r^2}\approx\frac{\alpha^2}{\sin^2\alpha r}.
\label{E19}
\end{equation}
By inserting approximation expression (\ref{E19}) and introducing a new variable of the form $z=\sin^2\alpha r$ into equation (\ref{E14}), we can easily find
\begin{equation}
\frac{d^2G_{n\kappa}(z)}{dz^2}+\frac{1-2z}{2z(1-z)}\frac{dG_{n\kappa}}{dz}-\left(\frac{\bar{\delta}\alpha^2+\bar{\gamma_1}}{4\alpha^2z^2(1-z)}+\frac{\bar{\gamma_2}}{4\alpha^2z(1-z)^2}+\frac{\bar{\beta}^2}{4\alpha^2z(1-z)}\right)G_{n\kappa}=0
\label{E20}
\end{equation}
Now we reach a position that the differential equation is appropriate for applying AIM. Therefore equation (\ref{E20}) should have a solution in the form of normalized wave functions which should respect the boundary conditions that is $G_{n\kappa}(0)=0$ at $z=0$ for $r\rightarrow\infty$ and $G_{n\kappa}(1)=0$ at $z=1$ for $r\rightarrow0$. As a result, we take the wave functions of the form
\begin{equation}
G_{n\kappa}(z)=z^{\bar{p}}(1-z)^{\bar{q}}g_{n\kappa}(z),
\label{E21}
\end{equation}
where we have intoduced the following notations for mathematical simplicity
\begin{eqnarray}
\bar{q}&=&\frac{1}{4}+\frac{1}{4}\sqrt{1+4\frac{\bar{\gamma_2}}{\alpha^2}}\nonumber\\
\bar{p}&=&\frac{1}{4}+\frac{1}{4}\sqrt{1+4\left(\bar{\delta}+\frac{\bar{\gamma_1}}{\alpha^2}\right)}
\label{E22}
\end{eqnarray}
Substituting equation (\ref{E21}) into equation (\ref{E20}) yields the following second order differential equation
\begin{equation}
g_{n\kappa}''(z)-\left[\frac{4z(\bar{p}+\bar{q}+1/2)-(4\bar{p}+1)}{2z(1-z)}\right]g_{n\kappa}'(z)-\left[\frac{\frac{\bar{\beta}^2}{\alpha^2}+4(\bar{p}+\bar{q})^2}{4z(1-z)}\right]g_{n\kappa}(z)=0,
\label{E23}
\end{equation}
which is suitable to an AIM solutions. Thus, by comparing equation (\ref{E23}) with equation (\ref{E1}), we can obtain $\lambda_0(z)$ and $s_0(z)$ and by using the recursion relation (\ref{E4}), we calculate $\lambda_k(z)$ and $s_k(z)$. Combining these results with the termination condition (\ref{E5}), we have the following  expressions for eigenvalues,
\begin{eqnarray}
s_0(z)\lambda_1(z)&=&s_1(z)\lambda_0(z)\ \ \ \ \ \ \ \Rightarrow\ \ \ \ \ \ \ \bar{\beta}^2=-4\alpha^2(\bar{p}+\bar{q})^2\nonumber\\
s_1(z)\lambda_2(z)&=&s_2(z)\lambda_1(z)\ \ \ \ \ \ \ \Rightarrow\ \ \ \ \ \ \ \bar{\beta}^2=-4\alpha^2(\bar{p}+\bar{q}+1)^2\\
s_2(z)\lambda_3(z)&=&s_2(z)\lambda_0(z)\ \ \ \ \ \ \ \Rightarrow\ \ \ \ \ \ \ \bar{\beta}^2=-4\alpha^2(\bar{p}+\bar{q}+2)^2\nonumber\\
\cdots etc.\nonumber
\label{E24}
\end{eqnarray}
If we generalize the above equations, the relativistic energy spectrum equation becomes
\begin{equation}
\bar{\beta}^2+4\alpha^2(\bar{p}+\bar{q}+n)^2=0.
\label{E25}
\end{equation}
By using the notations in equations (\ref{E18}) and (\ref{E22}), we obtain a more explicit expression for the relativistic energy spectrum equation as
\begin{equation}
(M+E_{n\kappa})(M-E_{n\kappa}+C_{ps})+4\alpha^2\left[n+\frac{1}{2}+\frac{1}{4}\left(\sqrt{1+\frac{4V_2}{\alpha^2}\left(E_{n\kappa}-M-C_{ps}\right)}+\bar{\eta}\right)\right]^2=0,
\label{E26}
\end{equation}
where
\begin{equation}
\bar{\eta}=\sqrt{1+4\bar{\delta}+\frac{4V_1}{\alpha^2}\left(E_{n\kappa}-M-C_{ps}\right)}.
\label{E27}
\end{equation}
Now, let us study the wave function of this system. Generally speaking, the differential equation we wish to solve should be transformed to the form $\cite{BJ5}$: 
\begin{equation}
y^{\prime \prime }(x)=2\left( \frac{\Lambda x^{N+1}}{1-bx^{N+2}}-\frac{m+1}{x%
}\right) y^{\prime }(x)-\frac{Wx^{N}}{1-bx^{N+2}}, 
\label{E28}
\end{equation}
where $a$, $b$ and $m$ are constants. The general solution of equation (\ref{E28}) is found as {\cite{BJ29}} 
\begin{equation}
y_{n}(x)=(-1)^{n}C_{2}(N+2)^{n}(\sigma )_{_{n}}{_{2}F_{1}(-n,t+n;\sigma;bx^{N+2})},  
\label{E29}
\end{equation}
where the following parameters have been used 
\begin{equation}
(\sigma )_{_{n}}=\frac{\Gamma {(\sigma +n)}}{\Gamma {(\sigma )}}\ \ ,\ \
\sigma =\frac{2m+N+3}{N+2}\ \ and\ \ \ t=\frac{(2m+1)b+2\Lambda }{(N+2)b}.
\label{E30}
\end{equation}
By comparing equations (\ref{E28}) with (\ref{E23}), we have $\Lambda =\bar{q}+1/4$, $b=1$, $N=-1$, $m=\bar{p}-3/4$ and $\sigma =2\bar{p}+1/2$. Thus the solution of equation (\ref{E24}) can easily be obtained as
\begin{equation}
G_{n\kappa}(z)=\bar{N}_{n\kappa}\frac{\Gamma(2\bar{p}+n+\frac{1}{2})}{\Gamma(2\bar{p}+\frac{1}{2})}z^{\bar{p}}(1-z)^{\bar{q}}\ {_2F_1}(-n, n+2(\bar{p}+\bar{q}); 2\bar{p}+\frac{1}{2};z),
\label{E31}
\end{equation}
where $\bar{N}_{n\kappa}$ is the normalization factor. In order to determine this factor, we need to express the solution equation (\ref{E31}) in terms of Jacobi polynomial \cite{BJ33}
\begin{equation}
G_{n\kappa}(z)=\bar{N}_{n\kappa}z^{\bar{p}}(1-z)^{\bar{q}}P_n^{(u,v)}(1-2z),
\label{E32}
\end{equation}
where
\begin{equation}
u=2\bar{p}-\frac{1}{2}, \ \ \ \ \ \ v=2\bar{q}-\frac{1}{2}.
\label{E33}
\end{equation}
Having done this, we can easily determine the normalization factor using the normalization condition $\int^1_{0}|G_{n\kappa}(z)|^2dz=1$, so that
\begin{equation}
\frac{\bar{N}_{n\kappa}^2}{2^{u+v+2}}\int^1_{-1}(1-x)^{u+\frac{1}{2}}(1+x)^{v+\frac{1}{2}}\left|P_n^{(u,v)}\right|^2dx=1, \ \ \ x=1-2z.
\label{E34}
\end{equation}
Now using the following relation of the hypergeometric function \cite{BJ33}
\begin{equation}
P_n^{(u,v)}(x)=\frac{1}{2^n}\sum^\infty_{m=0}\left(\begin{array}{lr}     
n+u\\      
\ \ m
\end{array}\right)\left(\begin{array}{lr}     
n+v\\      
n-m
\end{array}\right)(x-1)^{n-m}(1+x)^m,
\label{E35}
\end{equation}
and integral \cite{BJ6}
\begin{eqnarray}
\int^1_{-1}(1-x)^\alpha(1+x)^\beta P_n^{u,v}(x)dx=\frac{2^{\alpha+\beta+1}\Gamma(\beta+1)\Gamma(\alpha+1)\Gamma(n+1+u)}{n!\Gamma(\alpha+\beta+2)\Gamma(1+\alpha)}\nonumber\\
\times\ {_3F_2}(-n,u+v+n+1,\alpha+1;\beta+1,u+1,\alpha+\beta+2;1),
\label{E36}
\end{eqnarray}
we obtain the normalization constant as
\begin{equation}
\bar{C_2}=\frac{1}{\sqrt{\bar{I}_n(u,v)}},
\label{E37}
\end{equation}
where
\begin{eqnarray}
\bar{I}_n(u,v)&=&\frac{1}{2}\sum^\infty_{m=0}(-1)^{n-m+1}\left(\begin{array}{lr}     
n+u\\      
\ \ m
\end{array}\right)\left(\begin{array}{lr}     
n+v\\      
n-m
\end{array}\right)\frac{\Gamma(n-m+u+\frac{3}{2})\Gamma(m+v+\frac{3}{2})\Gamma(n+1+u)}{2n!\Gamma(n+u+v+3)\Gamma(1+u)}\nonumber\\
&\times&\ {_3F_2}(-n,u+v+n+1,n-m+u+\frac{3}{2};m+v+\frac{3}{2},u+1,n+u+v+3;1).
\label{E38}
\end{eqnarray}
The upper component spinor of the Dirac equation for the tPT potential plus a Coulomb-like tensor potential can be calculated from the relation
\begin{equation}
F_{n\kappa}(r)=\frac{1}{M-E_{n\kappa}+C_{ps}}\left[\frac{d}{dr}-\frac{\kappa}{r}+U(r)\right]G_{n\kappa}(r).
\label{E39}
\end{equation}
\subsection{Approximate bound-states for the spin symmetric limit}
In this symmetry, we obtain the solution of equation (\ref{E17}) by using a transformation of the form $z=\sin^2\alpha r$ so that equation (\ref{E17}) can be recast into a similar form of equation (\ref{E20})
\begin{equation}
\frac{d^2F_{n\kappa}(z)}{dz^2}+\frac{1-2z}{2z(1-z)}\frac{dF_{n\kappa}(z)}{dz}-\left(\frac{{\delta}\alpha^2+\gamma_1}{4\alpha^2z^2(1-z)}+\frac{\gamma_2}{4\alpha^2z(1-z)^2}+\frac{\beta}{4\alpha^2z(1-z)}\right)F_{n\kappa}(z)=0.
\label{E40}
\end{equation}
It is clearly seen that equation (\ref{E40}) is identical with the equation (\ref{E20}), therefore, the relativistic energy spectrum equation is obtained as
\begin{equation}
(M-E_{n\kappa})(M+E-C_{ps})+4\alpha^2\left[n+\frac{1}{2}+\frac{1}{4}\left(\eta+\sqrt{1+\frac{4}{\alpha^2}\left(E+M-C_{s}\right)V_2}\right)\right]^2=0,
\label{E41}
\end{equation}
where
\begin{equation}
\eta=\sqrt{1+4\delta+\frac{4\gamma_1}{\alpha^2}}.
\label{E42}
\end{equation}
The corresponding lower component spinor $F_{nk}(z)$ is obtained as
\begin{eqnarray}
F_{n\kappa}(z)&=&N_{n\kappa}\frac{\Gamma(2p+n+\frac{1}{2})}{\Gamma(2p+\frac{1}{2})}z^{p}(1-z)^{q}\ {_2F_1}(-n, n+2(p+q); 2p+\frac{1}{2};z)\nonumber\\
&=&C_2z^{p}(1-z)^{q}P_n^{(c,d)}(1-2z),
\label{E43}
\end{eqnarray}
where
\begin{eqnarray}
c&=&2p-\frac{1}{2}=\frac{1}{2}\sqrt{1+4\left(\delta+\frac{\gamma_1}{\alpha^2}\right)},\nonumber\\
d&=&2q-\frac{1}{2}=\frac{1}{2}\sqrt{1+16p+4\gamma_2}.
\label{E44}
\end{eqnarray}
The normalization constant $N_{n\kappa}$ can easily be obtained as
\begin{equation}
N_{n\kappa}=\frac{1}{\sqrt{I_n(c,d)}},
\label{E45}
\end{equation}
where
\begin{eqnarray}
I_n(c,d)&=&\frac{1}{2}\sum^\infty_{m=0}(-1)^{n-m+1}\left(\begin{array}{lr}     
n+c\\      
\ \ m
\end{array}\right)\left(\begin{array}{lr}     
n+d\\      
n-m
\end{array}\right)\frac{\Gamma(n-m+c+\frac{3}{2})\Gamma(m+d+\frac{3}{2})\Gamma(n+1+c)}{2n!\Gamma(n+c+d+3)\Gamma(1+c)}\nonumber\\
&\times&\ {_3F_2}(-n,c+d+n+1,n-m+c+\frac{3}{2};m+d+\frac{3}{2},c+1,n+c+d+3;1).
\label{E47}
\end{eqnarray}
Again, the lower component spinor of the Dirac equation for the tPT potential plus a Coulomb-like tensor potential can be calculated from the relation
\begin{equation}
G_{n\kappa}(r)=\frac{1}{M+E_{n\kappa}-C_{s}}\left[\frac{d}{dr}-\frac{\kappa+A}{r}\right]F_{n\kappa}(r).
\label{E48}
\end{equation}
Now, let us discuss the non-relativistic limit of the energy eigenvalues of our solution. If we take $C_s=0$ and put $S(r)=V(r)=\Sigma(r)$, the non-relativistic limit of energy equation (\ref{E41}) under the following appropriate transformations $M+E_{n\kappa}\rightarrow 2\mu$ and $M-E_{n\kappa}\rightarrow -E_{n\ell}$ become
\begin{equation}
E_{n\ell}=\frac{\alpha^2}{8\mu}\left[2+\sqrt{1+\frac{8\mu V_2}{\alpha^2}}+\sqrt{(1+2\ell)^2+\frac{8\mu V_1}{\alpha^2}}+4n\right]^2.
\label{E50}
\end{equation}
This is found in excellent agreement with \cite{BJ37,BJ38} and when $\ell=0$ it is identical to \cite{BJ32}.
\section{Numerical Results}
\label{sec5}
In table \ref{tab1}, in view of p-spin symmetry, for the given parameter values and when $C_{ps}=-5.0fm^{-1}$ we noticed the degenerate doublets ($ns_{1/2}$, $(n-1)d_{3/2}$), ($np_{3/2}$, $(n-1)f_{5/2}$), ($nd_{5/2}$, $(n-1)g_{7/2}$)

\begin{table}[!t]
\caption{ The bound state energy eigenvalues $(fm^{-1})$ of the p-spin symmetry tPT potential for various values of $n$ and $\protect\kappa $ with $M=1.0 fm^{-1}$, $C_{ps}=-5.0 fm^{-1}$, $V_1=-0.002fm^{-1}$, $V_2=0.003fm^{-1}$ and $\alpha=0.01$.}
\label{tab1}\vspace*{15pt} {\scriptsize
\begin{tabular}{|c|c|c|c|c|c|c|}\hline
$\tilde{\ell}$ & $n$ & $\kappa $ & $(\ell ,j)$ & $E_{n\kappa }(fm^{-1})(A=0)$ & $%
E_{n\kappa }(fm^{-1})(A=0.5)$ & $E_{n\kappa }(fm^{-1})(A=1)$  \\[2.5ex] \hline
1 & 1 & -1 & $1s_{\frac{1}{2}}$ 	&-4.00084675171	&-4.00074396128	&-0.99652927726	 \\[1ex]
&  &  & 	&-0.99651749280	&-0.99652483503	&-4.00062839464	\\[1ex]\hline
2 & 1 & -2 & $1p_{\frac{3}{2}}$ 	&-4.00103061187	&-0.99650733888	&-0.99651749280	 \\[1ex]
&  &  & 	&-0.99649448980	&-4.00094153754	&-4.00084675171	\\[1ex]\hline
3 & 1 & -3 & $1d_{\frac{5}{2}}$  	&-4.00119658319	&-0.99647908416	&-4.00103061187	 \\[1ex]
&  &  & 	&-0.99646127635	&-4.00111533113	&-0.99649448980	\\[1ex]\hline
4 & 1 & -4 & $1f_{\frac{7}{2}}$ 	&-0.99641911259	&-0.99644122998	&-4.00119658319	 \\[1ex]
&  &  &  	&-4.00135100100	&-4.00127498841	&-0.99646127635	\\[1ex]\hline
1 & 2 & -1 & $2s_{\frac{1}{2}}$ 	&-4.00097957046	&-4.00087663841	&-0.99639619042	 \\[1ex]
&  &  &  	&-0.99638440732	&-0.99639174872	&-4.00076063954	\\[1ex]\hline
2 & 2 & -2 & $2p_{\frac{3}{2}}$ 	&-4.00116354206	&-0.99637425458	&-4.00097957046	 \\[1ex]
&  &  &	&-0.99636140695	&-4.00107442646	&-0.99638440732	\\[1ex]\hline
3 & 2 & -3 & $2d_{\frac{5}{2}}$ 	&-0.99632819728	&-0.99634600310	&-4.00116354206	 \\[1ex]
&  &  &  	&-4.00132955810	&-4.00124828793	&-0.99636140695	\\[1ex]\hline
4 & 2 & -4 & $2f_{\frac{7}{2}}$ 	&-0.99628603824	&-4.00140797604	&-4.00132955810	 \\[1ex]
&  &  &	&-4.00148399768	&-0.99630815318	&-0.99632819728	\\[1ex]\hline
1 & 1 & 2 & $0d_{\frac{3}{2}}$	&-0.99651749280	&-0.99650733888	&-0.99649448980	 \\[1ex]
&  &  & 	&-4.00084675171	&-4.00094153754	&-4.00103061187	\\[1ex]\hline
2 & 1 & 3 & $0f_{\frac{5}{2}}$ 	&-4.00103061187	&-0.99647908416	&-0.99646127635	 \\[1ex]
&  &  & 	&-0.99649448980	&-4.00111533113	&-4.00119658319	\\[1ex]\hline
3 & 1 & 4 & $0g_{\frac{7}{2}}$ 	&-0.99646127635	&-0.99644122998	&-4.00135100100	 \\[1ex]
&  &  & 	&-4.00119658319	&-4.00127498841	&-0.99641911259	\\[1ex]\hline
4 & 1 & 5 & $0h_{\frac{9}{2}}$ 	&-0.99641911259	&-4.00142496554	&-4.00149715088	 \\[1ex]
&  &  & 	&-4.00135100100	&-0.99639509068	&-0.99636932608	\\[1ex]\hline
1 & 2 & 2 & $1d_{\frac{3}{2}}$	&-0.99638440732	&-4.00107442646	&-0.99636140695	 \\[1ex]
&  &  &	&-4.00097957046	&-0.99637425458	&-4.00116354206	\\[1ex]\hline
2 & 2 & 3 & $1f_{\frac{5}{2}}$ 	&-4.00116354206	&-4.00124828793	&-4.00132955810	 \\[1ex]
&  &  &	&-0.99636140695	&-0.99634600310	&-0.99632819728	\\[1ex]\hline
3 & 2 & 4 & $1g_{\frac{7}{2}}$ 	&-4.00132955810	&-4.00140797604	&-0.99628603824	 \\[1ex]
&  &  &  	&-0.99632819728	&-0.99630815318	&-4.00148399768	\\[1ex]\hline
4 & 2 & 5 & $1h_{\frac{9}{2}}$ 	&-4.00148399768	&-0.99626201898	&-0.99623625721	 \\[1ex]
&  &  &  	&-0.99628603824	&-4.00155796870	&-4.00163015861	\\[1ex]\hline\hline
\end{tabular}%
} \vspace*{-1pt}
\end{table}

\begin{table}[!t]
\caption{ The bound state energy eigenvalues $(fm^{-1})$ of the p-spin symmetry tPT potential for various values of $n$ and $\protect\kappa $ with $M=1.0 fm^{-1}$, $C_{ps}=0$, $V_1=-0.002fm^{-1}$, $V_2=0.003fm^{-1}$ and $\alpha=0.01$.}
\label{tab2}\vspace*{15pt} {\scriptsize
\begin{tabular}{|c|c|c|c|c|c|c|}\hline\hline
$\tilde{\ell}$ & $n$ & $\kappa $ & $(\ell ,j)$ & $E_{n\kappa }(fm^{-1})(A=0)$ & $%
E_{n\kappa }(fm^{-1})(A=0.5)$ & $E_{n\kappa }(fm^{-1})(A=1)$  \\[2.5ex] \hline
1 & 1 & -1 & $1s_{\frac{1}{2}}$ 	&1.00128001185	&1.00112781103	&1.00096022670\\[2.5ex]\hline
2 & 1 & -2 & $1p_{\frac{3}{2}}$ 	&1.00155412543	&1.00142116103	&1.00128001185\\[2.5ex]\hline
3 & 1 & -3 & $1d_{\frac{5}{2}}$  	&1.00180227570	&1.00168074734	&1.00155412543\\[2.5ex]\hline
4 & 1 & -4 & $1f_{\frac{7}{2}}$ 	&1.00203337671	&1.00191959944	&1.00180227570\\[2.5ex]\hline
1 & 2 & -1 & $2s_{\frac{1}{2}}$ 	&1.00148070407	&1.00132886223	&1.00116221307\\[2.5ex]\hline
2 & 2 & -2 & $2p_{\frac{3}{2}}$ 	&1.00175447962	&1.00162165120	&1.00148070407\\[2.5ex]\hline
3 & 2 & -3 & $2d_{\frac{5}{2}}$ 	&1.00200244763	&1.00188100010	&1.00175447962\\[2.5ex]\hline
4 & 2 & -4 & $2f_{\frac{7}{2}}$ 	&1.00223342319	&1.00211970399	&1.00200244763\\[2.5ex]\hline
1 & 1 & 2 & $0d_{\frac{3}{2}}$	&1.00128001185	&1.00142116103	&1.00155412543\\[2.5ex]\hline
2 & 1 & 3 & $0f_{\frac{5}{2}}$ 	&1.00155412543	&1.00168074734	&1.00180227570\\[2.5ex]\hline
3 & 1 & 4 & $0g_{\frac{7}{2}}$ 	&1.00180227570	&1.00191959944	&1.00203337671\\[2.5ex]\hline
4 & 1 & 5 & $0h_{\frac{9}{2}}$ 	&1.00203337671	&1.00214411048	&1.00225219519\\[2.5ex]\hline
1 & 2 & 2 & $1d_{\frac{3}{2}}$	&1.00148070407	&1.00162165120	&1.00175447962\\[2.5ex]\hline
2 & 2 & 3 & $1f_{\frac{5}{2}}$ 	&1.00175447962	&1.00188100010	&1.00200244763\\[2.5ex]\hline
3 & 2 & 4 & $1g_{\frac{7}{2}}$ 	&1.00200244763	&1.00211970399	&1.00223342319\\[2.5ex]\hline
4 & 2 & 5 & $1h_{\frac{9}{2}}$ 	&1.00223342319	&1.00234410569	&1.00245214426\\[2.5ex]\hline
\hline
\end{tabular}%
} \vspace*{-1pt}
\end{table}

\begin{table}[!t]
\caption{ The bound state energy eigenvalues $(fm^{-1})$ of the spin symmetry tPT potential for various values of $n$ and $\protect\kappa $ with $M=1.0 fm^{-1}$, $C_{s}=5.0 fm^{-1}$, $V_1=0.002fm^{-1}$, $V_2=-0.003fm^{-1}$ and $\alpha=0.01$.}
\label{tab3}\vspace*{15pt} {\scriptsize
\begin{tabular}{|c|c|c|c|c|c|c|}\hline\hline
$n$& $\tilde{\ell}$ & $\kappa $ & $(\ell ,j)$ & $E_{n\kappa }(fm^{-1})(A=0)$ & $%
E_{n\kappa }(fm^{-1})(A=0.5)$ & $E_{n\kappa }(fm^{-1})(A=1)$  \\[2.5ex] \hline
0	&0	&-1	&0$s_{\frac{1}{2}}$	&0.99665059004	&4.00061127157	&0.99666237586	 \\[1ex]	
	&	&	&	&4.00071392089	&0.99665793312	&4.00049613269	\\[1ex]\hline	
1	&0	&-1	&$1s_{\frac{1}{2}}$	&4.00084675171	&4.00074396128	&4.00062839464	 \\[1ex]	
	&	&	&	&0.99651749280	&0.99652483503	&0.99652927726	\\[1ex]\hline	
2	&0	&-1	&$2s_{\frac{1}{2}}$	&4.00097957046	&4.00087663841	&0.99639619042	 \\[1ex]	
	&	&	&	&0.99638440732	&0.99639174872	&4.00076063954	\\[1ex]\hline	
3	&0	&-1	&$3s_{\frac{1}{2}}$	&4.00111237713	&0.99625867420	&4.00089286732	 \\[1ex]	
	&	&	&	&0.99625133365	&4.00100930296	&0.99626311539	\\[1ex]\hline	
0	&1	&-2	&$0p_{\frac{3}{2}}$	&4.00089766980	&4.00080863669	&4.00071392089	 \\[1ex]	
	&	&	&	&0.99662758439	&0.99664043494	&0.99665059004	\\[1ex]\hline	
\end{tabular}%
} \vspace*{-1pt}
\end{table}

\begin{table}[!t]
\begin{flushleft}
{Table \ref{tab3}. Continued}
\end{flushleft}\vspace*{-15pt} {\scriptsize 
\begin{tabular}[t]{|c|c|c|c|c|c|c|}
\hline
$n$ & $\tilde{\ell}$ & $\kappa $ & $(\ell ,j)$ & $E_{n\kappa }(fm^{-1})(A=0)$ & $%
E_{n\kappa }(fm^{-1})(A=0.5)$ & $E_{n\kappa }(fm^{-1})(A=1)$  \\[2.5ex] \hline
1	&1	&-2	&$1p_{\frac{3}{2}}$	&0.99649448980	&0.99650733888	&0.99651749280	 \\[1ex]	
	&	&	&	&4.00103061187	&4.00094153754	&4.00084675171	\\[1ex]\hline	
2	&1	&-2	&$2p_{\frac{3}{2}}$	&0.99636140695	&0.99637425458	&4.00097957046	 \\[1ex]	
	&	&	&	&4.00116354206	&4.00107442646	&0.99638440732	\\[1ex]\hline	
3	&1	&-2	&$3p_{\frac{3}{2}}$	&4.00129646038	&0.99624118210	&4.00111237713	 \\[1ex]	
	&	&	&	&0.99622833594	&4.00120730344	&0.99625133365	\\[1ex]\hline	
0	&2	&-3	&$0d_{\frac{5}{2}}$	&4.00106359645	&4.00098236248	&4.00089766980	 \\[1ex]	
	&	&	&	&0.99659436718	&0.99661217702	&0.99662758439	\\[1ex]\hline	
1	&2	&-3	&$1d_{\frac{5}{2}}$	&0.99646127635	&0.99647908416	&0.99649448980	 \\[1ex]	
	&	&	&	&4.00119658319	&4.00111533113	&4.00103061187	\\[1ex]\hline	
2	&2	&-3	&$2d_{\frac{5}{2}}$	&0.99632819728	&4.00124828793	&4.00116354206	 \\[1ex]	
	&	&	&	&4.00132955810	&0.99634600310	&0.99636140695	\\[1ex]\hline	
3	&2	&-3	&$3d_{\frac{5}{2}}$	&4.00146252118	&0.99621293382	&4.00129646038	 \\[1ex]	
	&	&	&	&0.99619513000	&4.00138123289	&0.99622833594	\\[1ex]\hline	
0	&3	&-4	&$0f_{\frac{7}{2}}$	&4.00121799251	&0.99657431858	&0.99659436718	 \\[1ex]	
	&	&	&	&0.99655219872	&4.00114198898	&4.00106359645	\\[1ex]\hline	
1	&3	&-4	&$1f_{\frac{7}{2}}$	&0.99641911259	&4.00127498841	&4.00119658319	 \\[1ex]	
	&	&	&	&4.00135100100	&0.99644122998	&0.99646127635	\\[1ex]\hline	
2	&3	&-4	&$2f_{\frac{7}{2}}$	&4.00148399768	&0.99630815318	&4.00132955810	 \\[1ex]	
	&	&	&	&0.99628603824	&4.00140797604	&0.99632819728	\\[1ex]\hline	
3	&3	&-4	&$3f_{\frac{7}{2}}$	&4.00161698256	&4.00154095184	&0.99619513000	 \\[1ex]	
	&	&	&	&0.99615297566	&0.99617508814	&4.00146252118	\\[1ex]\hline	
0	&1	&1	&$0p_{\frac{1}{2}}$	&0.99666237586	&0.99665793312	&4.00071392089	 \\[1ex]	
	&	&	&	&4.00049613269	&4.00061127157	&0.99665059004	\\[1ex]\hline	
1	&1	&1	&$1p_{\frac{1}{2}}$	&4.00062839464	&4.00074396128	&4.00084675171	 \\[1ex]	
	&	&	&	&0.99652927726	&0.99652483503	&0.99651749280	\\[1ex]\hline	
2	&1	&1	&$2p_{\frac{1}{2}}$	&4.00076063954	&0.99639174872	&4.00097957046	 \\[1ex]	
	&	&	&	&0.99639619042	&4.00087663841	&0.99638440732	\\[1ex]\hline	
3	&1	&1	&$3p_{\frac{1}{2}}$	&0.99626311539	&4.00100930296	&0.99625133365	 \\[1ex]	
	&	&	&	&4.00089286732	&0.99625867420	&4.00111237713	\\[1ex]\hline	
0	&2	&2	&$0d_{\frac{3}{2}}$	&4.00071392089	&4.00080863669	&4.00089766980	 \\[1ex]	
	&	&	&	&0.99665059004	&0.99664043494	&0.99662758439	\\[1ex]\hline	
1	&2	&2	&$1d_{\frac{3}{2}}$	&4.00084675171	&4.00094153754	&0.99649448980	 \\[1ex]	
	&	&	&	&0.99651749280	&0.99650733888	&4.00103061187	\\[1ex]\hline	
2	&2	&2	&$2d_{\frac{3}{2}}$	&4.00097957046	&0.99637425458	&0.99636140695	 \\[1ex]	
	&	&	&	&0.99638440732	&4.00107442646	&4.00116354206	\\[1ex]\hline	
3	&2	&2	&$3d_{\frac{3}{2}}$	&4.00111237713	&4.00120730344	&4.00129646038	 \\[1ex]	
	&	&	&	&0.99625133365	&0.99624118210	&0.99622833594	\\[1ex]\hline	
0	&3	&3	&$0f_{\frac{5}{2}}$	&4.00089766980	&0.99661217702	&4.00106359645	 \\[1ex]	
	&	&	&	&0.99662758439	&4.00098236248	&0.99659436718	\\[1ex]\hline	
1	&3	&3	&$1f_{\frac{5}{2}}$	&0.99649448980	&4.00111533113	&4.00119658319	 \\[1ex]	
	&	&	&	&4.00103061187	&0.99647908416	&0.99646127635	\\[1ex]\hline	
2	&3	&3	&$2f_{\frac{5}{2}}$	&0.99636140695	&0.99634600310	&0.99632819728	 \\[1ex]	
	&	&	&	&4.00116354206	&4.00124828793	&4.00132955810	\\[1ex]\hline	
3	&3	&3	&$3f_{\frac{5}{2}}$	&4.00129646038	&0.99621293382	&4.00146252118	\\[1ex]	
	&	&	&	&0.99622833594	&4.00138123289	&0.99619513000	\\[1ex]\hline\hline
\end{tabular}
}\label{tab5b} \vspace*{-1pt}
\end{table}

\begin{table}[!t]
\caption{ The bound state energy eigenvalues $(fm^{-1})$ of the spin symmetry tPT potential for various values of $n$ and $\protect\kappa $ with $M=1.0 fm^{-1}$, $C_{s}=0 fm^{-1}$, $V_1=0.002fm^{-1}$, $V_2=-0.003fm^{-1}$ and $\alpha=0.01$.}
\label{tab4}\vspace*{15pt} {\scriptsize
\begin{tabular}{|c|c|c|c|c|c|c|}\hline\hline
$\tilde{\ell}$ & $n$ & $\kappa $ & $(\ell ,j)$ & $E_{n\kappa }(fm^{-1})(A=0)$ & $%
E_{n\kappa }(fm^{-1})(A=0.5)$ & $E_{n\kappa }(fm^{-1})(A=1)$  \\[2.5ex] \hline
0	&0	&-1	&0$s_{\frac{1}{2}}$	&-1.00107927801	&-1.00092671625	&-1.00075818292		 \\[2.5ex]\hline
1	&0	&-1	&$1s_{\frac{1}{2}}$	&-1.00128001185	&-1.00112781103	&-1.00096022670		 \\[2.5ex]\hline
2	&0	&-1	&$2s_{\frac{1}{2}}$	&-1.00148070407	&-1.00132886223	&-1.00116221307		 \\[2.5ex]\hline
3	&0	&-1	&$3s_{\frac{1}{2}}$	&-1.00168135469	&-1.00152986990	&-1.00136414241		 \\[2.5ex]\hline
0	&1	&-2	&$0p_{\frac{3}{2}}$	&-1.00135373059	&-1.00122062989	&-1.00107927801		 \\[2.5ex]\hline
1	&1	&-2	&$1p_{\frac{3}{2}}$	&-1.00155412543	&-1.00142116103	&-1.00128001185		 \\[2.5ex]\hline
2	&1	&-2	&$2p_{\frac{3}{2}}$	&-1.00175447962	&-1.00162165120	&-1.00148070407		 \\[2.5ex]\hline
3	&1	&-2	&$3p_{\frac{3}{2}}$	&-1.00195479317	&-1.00182210043	&-1.00168135469		 \\[2.5ex]\hline
0	&2	&-3	&$0d_{\frac{5}{2}}$	&-1.00160206341	&-1.00148045411	&-1.00135373059		 \\[2.5ex]\hline
1	&2	&-3	&$1d_{\frac{5}{2}}$	&-1.00180227570	&-1.00168074734	&-1.00155412543		 \\[2.5ex]\hline
2	&2	&-3	&$2d_{\frac{5}{2}}$	&-1.00200244763	&-1.00188100010	&-1.00175447962		 \\[2.5ex]\hline
3	&2	&-3	&$3d_{\frac{5}{2}}$	&-1.00220257924	&-1.00208121240	&-1.00195479317		 \\[2.5ex]\hline
0	&3	&-4	&$0f_{\frac{7}{2}}$	&-1.00183329004	&-1.00171945463	&-1.00160206341		 \\[2.5ex]\hline
1	&3	&-4	&$1f_{\frac{7}{2}}$	&-1.00203337671	&-1.00191959944	&-1.00180227570		 \\[2.5ex]\hline
2	&3	&-4	&$2f_{\frac{7}{2}}$	&-1.00223342319	&-1.00211970399	&-1.00200244763		 \\[2.5ex]\hline
3	&3	&-4	&$3f_{\frac{7}{2}}$	&-1.00243342951	&-1.00231976829	&-1.00220257924		 \\[2.5ex]\hline
0	&1	&1	&$0p_{\frac{1}{2}}$	&-1.00075818292	&-1.00092671625	&-1.00107927801		 \\[2.5ex]\hline
1	&1	&1	&$1p_{\frac{1}{2}}$	&-1.00096022670	&-1.00112781103	&-1.00128001185		 \\[2.5ex]\hline
2	&1	&1	&$2p_{\frac{1}{2}}$	&-1.00116221307	&-1.00132886223	&-1.00148070407		 \\[2.5ex]\hline
3	&1	&1	&$3p_{\frac{1}{2}}$	&-1.00136414241	&-1.00152986990	&-1.00168135469		 \\[2.5ex]\hline
0	&2	&2	&$0d_{\frac{3}{2}}$	&-1.00107927801	&-1.00122062989	&-1.00135373059		 \\[2.5ex]\hline
1	&2	&2	&$1d_{\frac{3}{2}}$	&-1.00128001185	&-1.00142116103	&-1.00155412543		 \\[2.5ex]\hline
2	&2	&2	&$2d_{\frac{3}{2}}$	&-1.00148070407	&-1.00162165120	&-1.00175447962		 \\[2.5ex]\hline
3	&2	&2	&$3d_{\frac{3}{2}}$	&-1.00168135469	&-1.00182210043	&-1.00195479317		 \\[2.5ex]\hline
0	&3	&3	&$0f_{\frac{5}{2}}$	&-1.00135373059	&-1.00148045411	&-1.00160206341		 \\[2.5ex]\hline
1	&3	&3	&$1f_{\frac{5}{2}}$	&-1.00155412543	&-1.00168074734	&-1.00180227570		 \\[2.5ex]\hline
2	&3	&3	&$2f_{\frac{5}{2}}$	&-1.00175447962	&-1.00188100010	&-1.00200244763		 \\[2.5ex]\hline
3	&3	&3	&$3f_{\frac{5}{2}}$	&-1.00195479317	&-1.00208121240	&-1.00220257924		 \\[2.5ex]\hline\hline
\end{tabular}
} \vspace*{-1pt}
\end{table}

\begin{table}[!t]
\caption{ The bound state energy eigenvalues $(fm^{-1})$ of the p-spin
symmetry tPT potential for various values of $n$ and $\kappa$ with $A=1.0 $, $V_1=-0.002fm^{-1}$, $V_2=0.003fm^{-1}$, $C_{ps}=-5.0 fm^{-1}$ and $\alpha=0.01$ for various $M$ values.}
\label{tab5}\vspace*{15pt} {\scriptsize 
\begin{tabular}{|c|c|c|c|c|c|}
\hline
\multicolumn{1}{|c|}{} & \multicolumn{5}{|c|}{$E_{n\kappa}(fm^{-1})$} \\
[1.5ex] \hline
{} & {} & {} & {} & {} & {} \\[-1.0ex] 
$M$ & $1s_{\frac{1}{2}}$ & $1p_{\frac{3}{2}}$ & $1d_{\frac{5}{2}}$ & $
1f_{\frac{7}{2}}$ & $2p_{\frac{3}{2}}$ \\[2.5ex] \hline
0.1	&-4.90039392640	&-4.90052995635	&-4.90064482152	&-0.09729461539	&-4.90061310055	\\[1ex]
	&-0.09732700200	&-0.09732146762	&-0.09731056761	&-4.90074856469	&-0.09723820785	\\[1ex]\hline
0.2	&-0.19726363906	&-4.80055294367	&-4.80067280728	&-0.19722900196	&-0.19717083955	 \\[1ex]
	&-4.80041096535	&-0.19725771395	&-0.19724605202	&-4.80078106094	&-4.80063969332	\\[1ex]\hline
0.3	&-4.70042954473	&-4.70057801533	&-0.29717685178	&-0.29715857470	&-4.70066869719	 \\[1ex]
	&-0.29719572965	&-0.29718936662	&-4.70070333204	&-4.70081650610	&-0.29709854959	\\[1ex]\hline
0.4	&-0.39712271208	&-0.39711585572	&-0.39710238108	&-0.39708272579	&-4.60070045558	 \\[1ex]
	&-4.60044988331	&-4.60060546823	&-4.60073675770	&-4.60085532055	&-0.39702072145	\\[1ex]\hline
0.5	&-4.50047224329	&-4.50063565847	&-0.49702194739	&-0.49700073559	&-0.49693662590	 \\[1ex]
	&-0.49704392346	&-0.49703650814	&-4.50077351833	&-4.50089800858	&-4.50073538040	\\[1ex]\hline
0.6	&-0.59695857526	&-4.40066901686	&-4.40081413912	&-0.59691174485	&-0.59684539381	 \\[1ex]
	&-4.40049694173	&-0.59695052254	&-0.59693472524	&-4.40094518034	&-4.40077396991	\\[1ex]\hline
0.7	&-4.30052436563	&-0.69685693638	&-4.30085926163	&-4.30099758120	&-4.30081683274	 \\[1ex]
	&-0.69686572091	&-4.30070606962	&-0.69683972144	&-0.69681471839	&-0.69674597988	\\[1ex]\hline
0.8	&-4.20055499237	&-0.79675458195	&-4.20090967790	&-0.79670839608	&-4.20086472012	 \\[1ex]
	&-0.79676421325	&-4.20074746629	&-0.79673572951	&-4.20105613138	&-0.79663711411	\\[1ex]\hline
0.9	&-4.10058941764	&-0.89664202692	&-4.10096637726	&-4.10112198039	&-4.10091857028	 \\[1ex]
	&-0.89665264655	&-4.10079401812	&-0.89662126706	&-0.89659122671	&-0.89651723662	\\[1ex]\hline
1	&-4.00062839464	&-4.00084675171	&-4.00103061187	&-4.00119658319	&-0.99638440732	 \\[1ex]
	&-0.99652927726	&-0.99651749280	&-0.99649448980	&-0.99646127635	&-4.00097957046	\\[1ex]\hline
1.1	&-3.90067288966	&-1.09637874016	&-1.09635307082	&-3.90128180875	&-3.90104924536	 \\[1ex]
	&-1.0963919127	&-3.90090698555	&-3.90110399015	&-1.09631609904	&-1.09623617950	\\[1ex]\hline
1.2	&-3.80072416324	&-3.80097644160	&-1.19619402672	&-3.80138009857	&-3.80112958615	 \\[1ex]
	&-1.19623775209	&-1.19622290459	&-3.80118861311	&-1.19615255070	&-1.19606941778	\\[1ex]\hline
1.3	&-4.70042954473	&-4.70057801533	&-0.29717685178	&-4.70081650610	&-4.70066869719	 \\[1ex]
	&-0.29719572965	&-0.29718936662	&-4.70070333204	&-0.29715857470	&-0.29709854959	\\[1ex]\hline
1.4	&-1.39586328048	&-3.60115301847	&-3.6014037953	&-3.60163005154	&-3.60133382545	 \\[1ex]
	&-3.60085435156	&-1.39584382808	&-1.39580617694	&-1.39575247698	&-1.39566256488	\\[1ex]\hline
1.5	&-1.49563155612	&-3.50126761921	&-1.49556506384	&-1.49550287629	&-1.49540957039	 \\[1ex]
	&-3.50093871986	&-1.49560885888	&-3.50154348618	&-3.50179232744	&-3.50146637118	\\[1ex]\hline
1.6	&-3.40104156115	&-1.59533185124	&-1.59528010388	&-3.40199044576	&-1.59511056581	 \\[1ex]
	&-1.59535877115	&-3.40140749459	&-3.40171402081	&-1.59520700999	&-3.40162813925	\\[1ex]\hline
1.7	&-3.30116968106	&-3.30158203270	&-1.69493666456	&-1.69484917754	&-1.69475027299	 \\[1ex]
	&-1.69503157659	&-1.69499899825	&-3.30192686849	&-3.30223773705	&-3.30182997821	\\[1ex]\hline
1.8	&-3.20133369067	&-3.20180592116	&-1.79451240641	&-3.20255506406	&-3.20208885552	 \\[1ex]
	&-1.79462977645	&-1.79458933368	&-3.20219997935	&-1.79440528715	&-1.79430542808	\\[1ex]\hline
1.9	&-3.10155109245	&-1.89406922289	&-3.10256310344	&-1.89383621716	&-1.89373860058	 \\[1ex]
	&-1.89412112087	&-3.10210349324	&-1.89397126174	&-3.10297700153	&-3.10243286936	\\[1ex]\hline
2	&-1.9934503596	&-3.00251817685	&-3.00306935812	&-3.00356526814	&-3.00291213384	\\[1ex]
	&-3.00185293961	&-1.99338070398	&-1.99325059061	&-1.99307357655	&-1.99298508396	\\[1ex]\hline	\hline
\end{tabular}
} \vspace*{-1pt}
\end{table}

\begin{table}[!t]
\caption{ The bound state energy eigenvalues $(fm^{-1})$ of the spin
symmetry tPT potential for various values of $n$ and $\kappa$ with $A=1.0 $, $V_1=0.002fm^{-1}$, $V_2=-0.003fm^{-1}$, $C_{s}=5.0 fm^{-1}$ and $\alpha=0.01$ for various $M$ values.}
\label{tab6}\vspace*{15pt} {\scriptsize 
\begin{tabular}{|c|c|c|c|c|c|}
\hline
\multicolumn{1}{|c|}{} & \multicolumn{5}{|c|}{$E_{n\kappa}(fm^{-1})$} \\
[1.5ex] \hline
{} & {} & {} & {} & {} & {} \\[-1.0ex] 
$M$ & $1s_{\frac{1}{2}}$ & $1p_{\frac{3}{2}}$ & $0p_{\frac{1}{2}}$ & $
2f_{\frac{7}{2}}$ & $1f_{\frac{5}{2}}$ \\[2.5ex] \hline
0.1	&4.90031100228	&0.09740473026	&0.09722730832	&0.09737787681	&4.90069624178	\\[1ex]
	&0.09741026489	&4.90044680919	&4.90072801156	&4.90066535215	&0.09715495100	\\[1ex]\hline
0.2	&0.19735051704	&0.19734459162	&4.80075960665	&0.19731587830	&0.19708396838	 \\[1ex]
	&4.80032445514	&4.80046619067	&0.19715917820	&4.80069423709	&4.80072643960	\\[1ex]\hline
0.3	&4.70033912433	&4.70048732963	&4.70079406804	&0.29724939385	&4.70075937520	 \\[1ex]
	&0.29728655078	&0.29728018736	&0.29708603541	&4.70072574339	&0.29700773632	\\[1ex]\hline
0.4	&0.39721785105	&4.60051047646	&0.39700724759	&4.60076024477	&0.39692559145	 \\[1ex]
	&4.60035518260	&0.39721099430	&4.60083180422	&0.39717786242	&4.60079543853	\\[1ex]\hline
0.5	&4.50037283704	&0.49713639534	&4.50087330516	&4.50079818954	&0.49683674861	 \\[1ex]
	&0.49714381118	&4.50053593143	&0.49692206612	&0.49710062052	&4.50083509722	\\[1ex]\hline
0.6	&4.40039233803	&4.40056405785	&0.59682959762	&0.59701687660	&4.40087891700	 \\[1ex]
	&0.59706371036	&0.59705565706	&4.40091916370	&4.40084012014	&0.59674027090	\\[1ex]\hline
0.7	&0.69697668496	&0.69696789975	&4.30097010397	&0.69692567839	&0.69663503020	 \\[1ex]
	&4.30041399122	&4.30059529950	&0.69672876634	&4.30088669926	&4.30092758886	\\[1ex]\hline
0.8	&0.79688169001	&0.79687205789	&0.79661826339	&4.20093874505	&0.79651965441	 \\[1ex]
	&4.20043817364	&4.20063020415	&4.20102701994	&0.79682586789	&4.20098196565	\\[1ex]\hline
0.9	&4.10046535582	&4.10066945600	&0.89649647884	&4.10099727956	&4.10104311249	 \\[1ex]
	&0.89677744766	&0.89676682692	&4.10109102824	&0.89671602164	&0.89639245604	\\[1ex]\hline
1	&4.00049613269	&4.00071392089	&4.00116354206	&0.99659436718	&4.00111237713	 \\[1ex]
	&0.99666237586	&0.99665059004	&0.99636140695	&4.00106359645	&0.99625133365	\\[1ex]\hline
1.1	&1.09653448962	&3.90076471090	&3.90124637660	&1.09645866601	&1.09609363330	 \\[1ex]
	&3.90053126759	&1.09652131538	&1.09621051352	&3.90113935763	&3.90119149033	\\[1ex]\hline
1.2	&1.19639125924	&3.80082327853	&3.80134190280	&3.80122673419	&3.80128271220	 \\[1ex]
	&3.80057175638	&1.19637640944	&1.19604054434	&1.19630604489	&1.19591594906	\\[1ex]\hline
1.3	&3.70061892325	&3.70089155767	&3.70145327762	&3.70132861751	&1.29571382942	 \\[1ex]
	&1.29622940344	&1.29621250199	&1.29584724308	&1.29613274660	&3.70138905128	\\[1ex]\hline
1.4	&3.60067456805	&1.39602512106	&3.60158479885	&3.60144894482	&3.60151460198	 \\[1ex]
	&1.39604457766	&3.60097218101	&1.39562492178	&1.39593375071	&1.39548133145	\\[1ex]\hline
1.5	&1.49583089015	&1.49580818698	&1.49536578661	&3.50159321853	&3.50166508269	 \\[1ex]
	&3.50074120092	&3.50106882676	&3.50174247115	&1.49570217736	&1.49521032146	\\[1ex]\hline
1.6	&1.59558011948	&3.40118679454	&3.40193494582	&1.59542831045	&1.59488933452	 \\[1ex]
	&3.40082243066	&1.59555319089	&1.59505883485	&3.40176936913	&3.40184872855	\\[1ex]\hline
1.7	&1.69528039211	&1.69524780054	&1.69468796424	&3.30198925857	&1.69450162464	 \\[1ex]
	&3.30092363917	&3.30133400857	&3.30217515729	&1.69509792051	&3.30207784517	\\[1ex]\hline
1.8	&3.20105322137	&1.79487335380	&3.20248334274	&1.79468921256	&1.79402163686	 \\[1ex]
	&1.79491381806	&3.20152286996	&1.79422854094	&3.20227145667	&3.20237167316	\\[1ex]\hline
1.9	&3.10122503324	&3.10177393271	&3.10289302854	&1.89416685828	&3.10276206133	 \\[1ex]
	&1.89445196156	&1.89440002611	&1.89364070856	&3.10264675123	&1.89340815892	\\[1ex]\hline
2	&3.00146368278	&1.99377663414	&1.99285510169	&3.00317013653	&3.00330577578	\\[1ex]
	&1.99384636185	&3.00212390413	&3.00346403560	&1.99346920195	&1.99258977338\\[1ex]\hline 
	\hline
\end{tabular}
} \vspace*{-1pt}
\end{table}

\begin{table}[!t]
\caption{ The bound state energy eigenvalues $(fm^{-1})$ of the p-spin
symmetry tPT potential for various values of $n$ and $\kappa$ with $A=1.0 $, $V_1=-0.002fm^{-1}$, $V_2=0.003fm^{-1}$, $M=1.0 fm^{-1}$ and $\alpha=0.01$ for various $C_{ps}$ values.}
\label{tab7}\vspace*{15pt} {\scriptsize 
\begin{tabular}{|c|c|c|c|c|c|}
\hline
\multicolumn{1}{|c|}{} & \multicolumn{5}{|c|}{$E_{n\kappa}(fm^{-1})$} \\
[1.5ex] \hline
{} & {} & {} & {} & {} & {} \\[-1.0ex] 
$C_{ps}$ & $1s_{\frac{1}{2}}$ & $1p_{\frac{3}{2}}$ & $1d_{\frac{5}{2}}$ & $
1f_{\frac{7}{2}}$ & $2p_{\frac{3}{2}}$ \\[2.5ex] \hline
	&-0.99921219950	&-0.99921205140	&-49.0000729061	&-49.0000749459	&-0.99920239080	\\[1ex]	
-40	&-49.0000395646	&-49.0000530984	&-0.99920342210	&-0.99921131220	&-49.0000929333	\\[1ex]\hline	
	&-44.0000441627	&-44.0000592712	&-0.99915625190	&-0.99916502910	&-44.0001037383	 \\[1ex]	
-35	&-0.99916608040	&-0.99916590460	&-44.0000813822	&-44.0000836594	&-0.99915503010	\\[1ex]\hline	
	&-0.99911091390	&-0.99911070110	&-0.99909975020	&-0.99910964120	&-0.99909827220	 \\[1ex]	
	&-39.0000499701	&-39.0000670680	&-39.0000920884	&-39.0000946656	&-39.0001173863	\\[1ex]\hline	
-30	&-34.0000575361	&-0.99904305190	&-0.99903040260	&-0.99904173360	&-0.99902856520	 \\[1ex]	
	&-0.99904331680	&-34.0000772268	&-34.0001060382	&-34.0001090063	&-34.0001351694	\\[1ex]\hline	
-25	&-29.0000678020	&-0.99895748920	&-29.0001249687	&-0.99895578880	&-0.99894015410	 \\[1ex]	
	&-0.99895783120	&-29.0000910124	&-0.99894252240	&-29.0001284677	&-29.0001593024	\\[1ex]\hline	
	&-0.99884494550	&-24.0001107891	&-24.0001521274	&-0.99884217290	&-24.0001939257	 \\[1ex]	
-20	&-24.0000825269	&-0.99884448100	&-0.99882616480	&-24.0001563884	&-0.99882295410	\\[1ex]\hline	
	&-19.0001054219	&-19.0001415468	&-0.99866193120	&-0.99868212840	&-19.0002477784	 \\[1ex]	
-15	&-0.99868619020	&-0.99868550860	&-19.0001943680	&-19.0001998154	&-0.99865723720	\\[1ex]\hline	
	&-0.99843932820	&-14.0001959457	&-0.99840516810	&-14.0002766319	&-0.99839738960	 \\[1ex]	
	&-14.0001458971	&-0.99843819160	&-14.0002690827	&-0.99843257320	&-14.0003430387	\\[1ex]\hline	
-10	&-9.00023681890	&-0.99797366094	&-0.99791882675	&-0.99796163488	&-0.99790228590	 \\[1ex]	
	&-0.99797611064	&-9.00031825471	&-9.00043709995	&-9.00044939125	&-9.00055728650	\\[1ex]\hline	
-5	&-0.99652927726	&-0.99651749280	&-0.99636140695	&-0.99646127635	&-4.00148399768	 \\[1ex]	
	&-4.00062839464	&-4.00084675171	&-4.00116354206	&-4.00119658319&-0.99628603824\\[1ex]\hline	
	\hline
\end{tabular}
} \vspace*{-1pt}
\end{table}

\begin{table}[!t]
\caption{ The bound state energy eigenvalues $(fm^{-1})$ of the spin
symmetry tPT potential for various values of $n$ and $\kappa$ with $A=1.0 $, $V_1=-0.002fm^{-1}$, $V_2=0.003fm^{-1}$, $M=1.0 fm^{-1}$ and $\alpha=0.01$ for various $C_{s}$ values.}
\label{tab8}\vspace*{15pt} {\scriptsize 
\begin{tabular}{|c|c|c|c|c|c|}
\hline
\multicolumn{1}{|c|}{} & \multicolumn{5}{|c|}{$E_{n\kappa}(fm^{-1})$} \\
[1.5ex] \hline
{} & {} & {} & {} & {} & {} \\[-1.0ex] 
$C_{s}$ & $0s_{\frac{1}{2}}$ & $0p_{\frac{3}{2}}$ & $2d_{\frac{5}{2}}$ & $
0f_{\frac{7}{2}}$ & $3p_{\frac{1}{2}}$ \\[2.5ex] \hline
5	&0.99652927726	&4.00084675171	&4.00071392089	&4.0013295581	&4.00132955810	\\[1ex]	
	&4.00062839464	&0.99651749280	&0.99665059004	&0.99632819728	&0.99632819728	\\[1ex]\hline	
10	&9.00023681890	&0.99797366094	&9.00026831929	&9.00049935066	&0.99791165490	 \\[1ex]	
	&0.99797611064	&9.00031825471	&0.99802364178	&0.99791165490	&9.00049935066	\\[1ex]\hline	
15	&0.99843932820	&14.0001959457	&0.99846895540	&0.99840180990	&14.0003073866	 \\[1ex]	
	&14.0001458971	&0.99843819160	&14.0001652004	&14.0003073866	&0.99840180990	\\[1ex]\hline	
20	&19.0001054219	&0.99868550860	&0.99870772840	&0.99865990880	&0.99865990880	 \\[1ex]	
	&0.99868619020	&19.0001415468	&19.0001193369	&19.0002220302	&19.0002220302	\\[1ex]\hline	
25	&24.0000825269	&0.99884448100	&24.0000934054	&0.99882478300	&24.0001737753	 \\[1ex]	
	&0.99884494550	&24.0001107891	&0.99886187090	&24.0001737753	&0.99882478300	\\[1ex]\hline	
30	&0.99895783120	&0.99895748920	&29.0000767317	&0.99894150390	&0.99894150390	 \\[1ex]	
	&29.0000678020	&29.0000910124	&0.99897177420	&29.0001427505	&29.0001427505	\\[1ex]\hline	
35	&0.99904331680	&0.99904305190	&0.99905517250-	&0.99902961290	&34.0001211254	 \\[1ex]	
	&34.0000575361	&34.0000772268	&34.0000651092	&34.0001211254	&0.99902961290	\\[1ex]\hline	
40	&39.0000499701	&0.99911070110	&0.99912122700	&0.99909911520	&39.0001051903	 \\[1ex]	
	&0.99911091390	&39.0000670680	&39.0000565444	&39.0001051903	&0.99909911520	\\[1ex]\hline	
45	&44.0000441627	&0.99916590460	&0.99917520690	&0.99915572700	&0.99915572700	 \\[1ex]	
	&0.99916608040	&44.0000592712	&44.0000499710	&44.0000929605	&44.0000929605	\\[1ex]\hline	
50	&0.99921219950	&49.0000530984	&0.99922038440	&49.0000832783	&49.0000832783	 \\[1ex]	
	&49.0000395646	&0.99921205140	&49.0000447668	&0.99920297920	&0.99920297920	\\[1ex]\hline	\hline	
\end{tabular}
} \vspace*{-1pt}
\end{table}
\newpage
\begin{figure*}
\centering
\includegraphics[height=85mm,width=150mm]{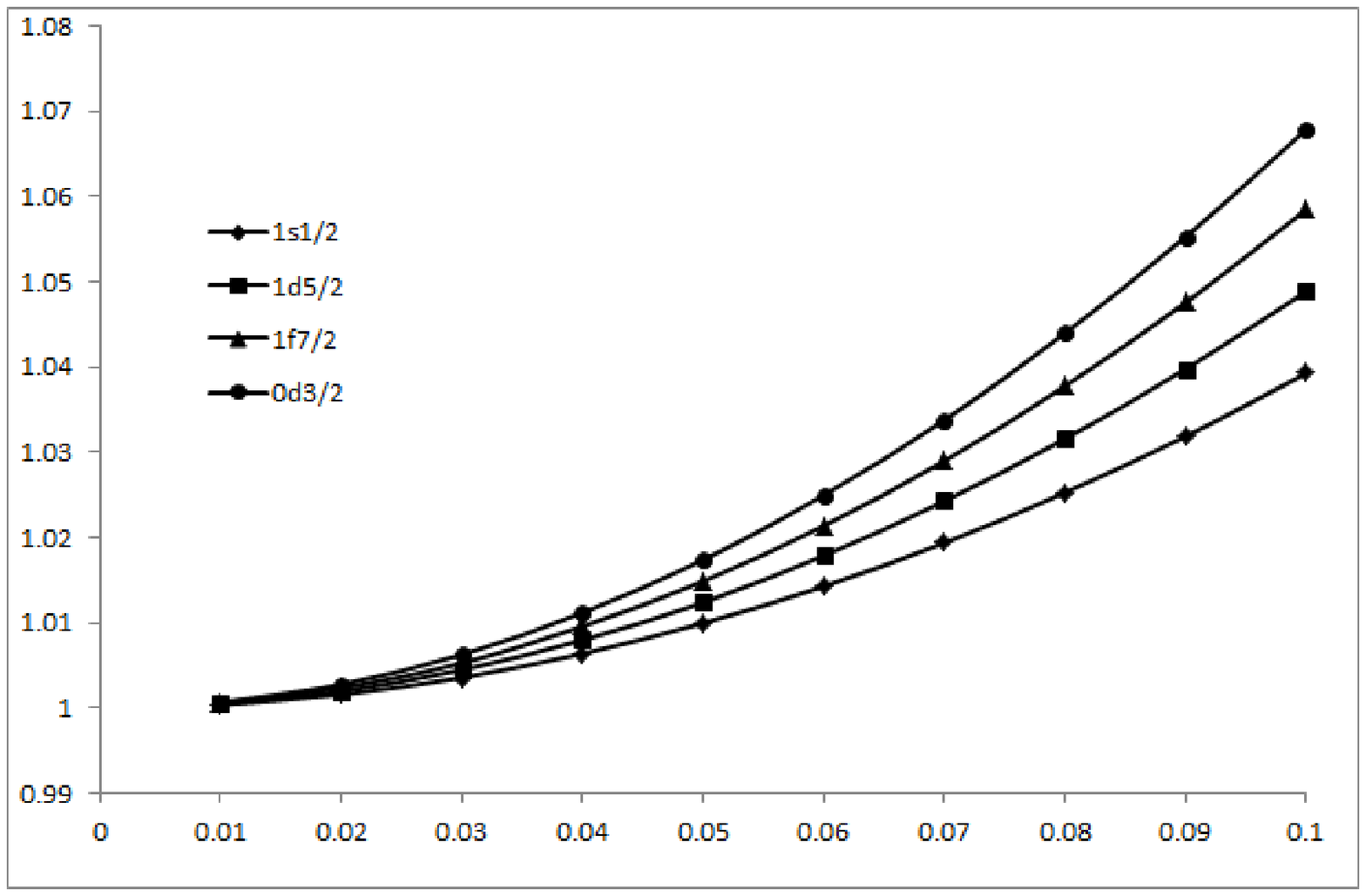}
\caption{\footnotesize The variation of energy as a function of the variable $\alpha$ for p-spin symmetry are displayed with parameters $M=1.0fm^{-1}$, $A=1.0$, $Cps=0$, $V_1=-0.002fm^{-1}$ and $V_2=0.003fm^{-1}$}
\label{fig:1}
\end{figure*}

\begin{figure*}
\centering
\includegraphics[height=85mm,width=150mm]{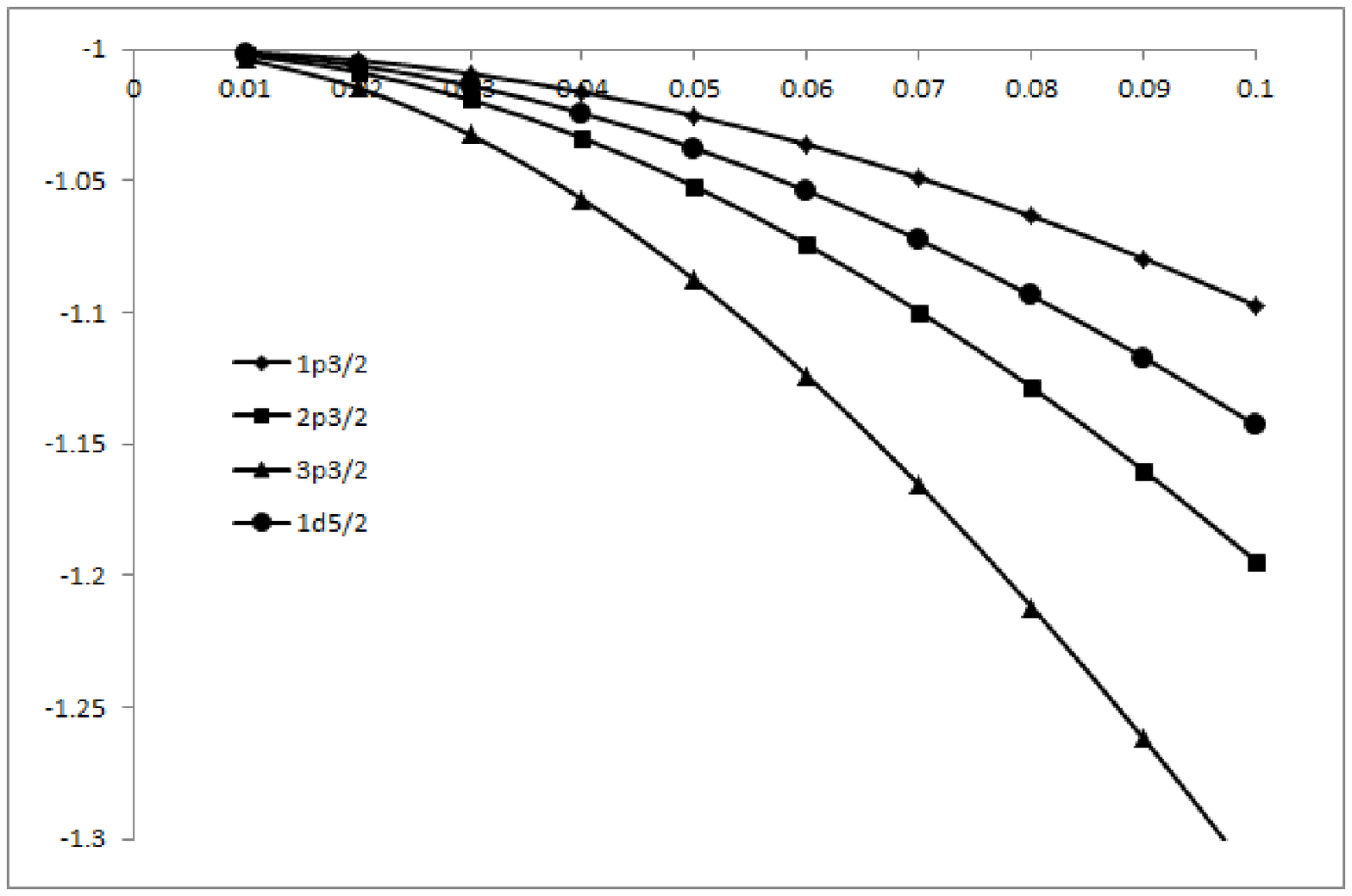}
\caption{\footnotesize The variation of energy as a function of the variable $\alpha$ for spin symmetry are displayed with parameters $M=1.0 fm^{-1}$, $A=1.0$, $Cs=0$, $V_1=0.002fm^{-1}$ and $V_2=-0.003fm^{-1}$}
\label{fig:2}
\end{figure*}
and ($nf_{7/2}$, $(n-1)h_{9/2}$) in the absence of tensor interaction $(A=0)$. Also, the energy levels are negative, i.e., the system becomes less attractive when the quantum numbers  increasing. In the presence of the tensor interaction, it removes the degeneracy in the above doublet states and new degenerate states appear as follows: ($1p_{3/2}$, $0d_{3/2}$), ($1d_{5/2}$, $0f_{5/2}$) and ($1f_{7/2}, 0g_{7/2}$). Further, increasing the strength of the tensor interaction as $A=1.0$, the degenerate doublets become ($1d_{5/2}$, $0d_{3/2}$) and ($1f_{7/2}$, $0f_{5/2}$). In table \ref{tab2}, when $C_{ps}=0$, the energy states become positive, i.e., the system is repulsive. 

In table \ref{tab3}, in view of spin symmetry, for the given parameter values and when $C_s=5fm^{-1}$ we noticed that the energy levels are repulsive and becoming weakly repulsive with increasing quantum numbers $n$ and $\kappa$ .The degenerate doublets are ($ns_{1/2}$, $nd_{3/2}$) and ($np_{3/2}$, $nf_{5/2}$)  in the absence of tensor interaction ($A=0$). In the presence of the tensor interaction, it removes the degeneracy in the above doublet states and new degenerate states appear as follows: ($ns_{1/2}$, $np_{1/2}$), ($np_{3/2}$, $nd_{3/2}$) and ($nd_{5/2}$, $nf_{5/2}$). Increasing the strength of the tensor interaction as $A=1.0$, the degenerate doublets become ($np_{3/2}$, $np_{1/2}$), ($nd_{5/2}$, $nd_{3/2}$)  and ($nf_{7/2}$, $nf_{5/2}$ ). In table \ref{tab4}, when $C_s=0$ (exact spin symmetry), the energy levels turn to negative and system becomes strongly attractive as the quantum numbers $n$ and $\kappa$ increasing.

In table \ref{tab5}, in presence of p-spin symmetry with parameter values $C_{ps}=-5.0fm^{-1}$ and $A=1.0$,  the energy levels are negative and subject to restriction $|E|<M$. The system becomes less attractive as quantum numbers increasing. In table \ref{tab6}, in presence of spin symmetry with parameter values $C_s=5.0fm^{-1}$ and $A=1.0$, the energy levels are positive and subject to restriction $|E|<M$ (the energy is less than mass but very near). The system becomes less repulsive with increasing quantum numbers. In table \ref{tab7}, when $A=1.0$ and $M=1.0fm^{-1}$, the system is weakly attractive when $C_{ps}$ increasing. The energy pushes up toward positive energy as $C_{ps}$ becoming less negative. In table \ref{tab8}, when $A=1.0$ and $M=1.0fm^{-1}$ the system is more repulsive when $C_{s}$  increasing. The energy pushes up toward positive energy as $C_{ps}$ becoming more positive.

\section{Concluding Remarks}
\label{sec6}
In this paper, we have studied the relativistic bound state solutions of a particle influenced by the scalar and vector tPT potentials including a Coulomb-like tensor interaction for any spin-orbit quantum number $\kappa$. We introduce a suitable approximation to substitute the spin-orbit centrifugal (p-centrifugal) coupling term to obtain approximate energy eigenvalue equation and the normalized two components of the radial wave functions. The spin and p-spin symmetric cases are studied for any  wave state. Our numerical results are presented in the presence/absence of tensor interaction and using various values of spin/p-spin constant for various   and   values. Further, the non-relativistic limit of our solution can be found by simply making an appropriate transformation.


\begin{thebibliography}{99}
\bibitem{BJ1} Bohr A, Hamamoto I and Mottelson B R 1982 Phys. Scr. {\bf26} 267 
\bibitem{BJ2} Dudek J, Nazarewicz W, Szymanski Z and Leander G A 1987 Phys. Rev. Lett. {\bf59} 1405
\bibitem{BJ3} Troltenier D, Bahri C and Draayer J P 1995 Nucl. Phys. A {\bf586} 53
\bibitem{BJ4} Page P R, Goldman T and Ginocchio J N 2001 Phys. Rev. Lett. {\bf86} 204
\bibitem{BJ5} Ginocchio J N, Leviatan A, Meng J and Zhou S G 2004 Phys. Rev. C {\bf69} 034303
\bibitem{BJ6} Lu Bing-Nan, Zhao En-Guang and Zhou Shan-Gui 2012 Phys. Rev. Lett. {\bf109} 072501
\bibitem{BJ7} Guo Jian-You 2012 Phys. Rev. C {\bf85} 021302(R)
\bibitem{BJ8} Chen Shou-Wan and Guo Jian-You 2012 Phys. Rev. C {\bf85} 054312
\bibitem{BJ9} Guo Jian-You, Zhou Fang, Guo Feng-Liang and Zhou Jian-Hong 2007 Int. J. Mod. Phys. A {\bf22} 4825
\bibitem{BJ10} Zhou Yan and Guo Jian-You 2008 Chin. Phys. B {\bf17} 380
\bibitem{BJ11} Ginocchio J N 1997 Phys. Rev. Lett. {\bf78} (3) 436
\bibitem{BJ12} Hecht K T and Adler A 1969 Nucl. Phys. A {\bf137} 129
\bibitem{BJ13} Arima A, Harvey M and Shimizu K 1969 Phys. Lett. B {\bf30} 517
\bibitem{BJ14} Zhou S G, Meng J and Ring P 2003 Phys. Rev. Lett. {\bf91} 262501
\bibitem{BJ15} He X T, Zhou S G, Meng J, Zhao E G and Scheid W 2006 Euro. Phys. J. A {\bf28} 265
\bibitem{BJ16} Moshinsky M and Szczepaniak A 1989 J. Phys. A: Math. Gen. {\bf22} L817
\bibitem{BJ17} Furnstahl R F, Rusnak J J and Serot B D 1998 Nucl. Phys. A {\bf632} 607
\bibitem{BJ18} Kukulin V I, Loyola G and Moshinsky M 1991 Phys. Lett. A {\bf158} 19
\bibitem{BJ19} Mao G 2003 Phys. Rev. C {\bf67} 044318
\bibitem{BJ20} Lisboa R, Malheiro M, de Castro A S, Alberto P and Fiolhais M 2004 Phys. Rev. C {\bf69} 0243319
\bibitem{BJ21} Alberto P, Lisboa R, Malheiro M and de Castro A S 2005 Phys. Rev. C {\bf71} 034313 
\bibitem{BJ22} Aktay H 2007 J. Phys. A: Math. Theor. {\bf40} 6427
\bibitem{BJ23} Akcay H 2009 Phys. Lett. A {\bf373} 616
\bibitem{BJ24} Ikhdair S M and Sever R 2010 Appl. Math. Comput. {\bf216} 545; Ikhdair S M and Sever R 2012 Appl. Math. Comput. {\bf218} 10082
\bibitem{BJ25} Hamzavi M, Ikhdair S M and Ita B I 2012 Phys. Scr. {\bf85} 045009; Eshghi M, Hamzavi M and Ikhdair S M 2012 Adv. High Energy Phys. {\bf2012} 873619
\bibitem{BJ26} Hamzavi M and Ikhdair S M 2010 Few-Body Syst. DOI : 10.1007/s00601-012-0475-2
\bibitem{BJ27} P\"oschl G and Teller E 1933 Z. Phys. {\bf83} 143 
\bibitem{BJ28} Chen G 2001 Acta Phys. Sinica {\bf50} 1651 (in Chinese)
\bibitem{BJ29} Zhang M C and Wang Z B 2006 Acta  Phys. Sinica {\bf55} 0525 (in Chinese)
\bibitem{BJ30} Liu X Y, Wei G F, Cao X W and Bai H G 2010 Int. J. Theor. Phys. {\bf49} 343
\bibitem{BJ31} Candemir N 2012 Int. J. Mod. Phys. E {\bf21} (6) 1250060
\bibitem{BJ32} Hamzavi M and Rajabi A A 2011 Int. J. Quant. Chem. {\bf112} 1592
\bibitem{BJ33} Hamzavi M, Ikhdair S M and Thylwe K E 2012 Int. J. Mod. Phys. E {\bf21} (12) 1250097 
\bibitem{BJ34} Ciftci H, Hall R L and Saad N 2003 J. Phys. A: Math Gen. {\bf36} 11807 
\bibitem{BJ35} Ciftci H, Hall R L and Saad N 2005 Phys. Lett. A: {\bf340} 388
\bibitem{BJ36}  Falaye B J 2012 \textit{Cent. Eur. J. Phys.} {\bf10(4)} 960\\
 Falaye B J 2012 \textit{Few Body Sys.} {\bf53} 557\\
 Falaye B J 2012 \textit{Few Body Sys.} {\bf53} 563\\
 Falaye B J 2012 \textit{J. Math. Phys.} {\bf53} 082107
\bibitem{BJ37} Falaye B J 2012 \textit{Can. J. Phys.} {\bf90}  1259\\
Corrigendum: Energy spectrum for trigonometric P\"{o}schl-Teller potential solved by the asymptotic iteration method; Can. J. Phys. 90 (2012) 1259 (to appear in Can. J. Phys. 2013)
\bibitem{BJ38} Hamzavi M and Ikhdair S M 2012 Mol. Phys. {\bf110(24)} 3031
\bibitem{N1} Hassanabadi H, Maghsoodi E, Zarrinkamar S and Rahimov H 2012 J. Math. Phys. {\bf53} 022104
\bibitem{N2} Hassanabadi H, Maghsoodi E, Zarrinkamar S and Rahimov H 2011, Modern Phys. Lett. A {\bf26} 2703 
\bibitem{N3} Maghsoodi E, Hassanabadi H and Zarrinkamar, S 2012  Few Body Syst. {\bf42} 2008
\bibitem{N4} Ikot A N 2012 Few-Body Syst. {\bf53} 549
\bibitem{N5} Oluwadare O J, Oyewumi K J, Akoshile C O and Babalola O A 2012 Phys. Scr. {\bf86} 035002
\bibitem{N6} Oyewumi K J and Akoshile C O 2010 Euro. Phys. J. A {\bf45} 311
\bibitem{N7} Ikhdair S M and Falaye B J 2013 Phys. Scr. 87  035002
\bibitem{BJ39} Gradshteyn I S and I. M. Ryzhik 1994, \textit{Tables of Integrals, Series and Products} (Academic Press: New York)
\bibitem{BJ40} Ikhdair S M 2012 Cent. Eur. J. Phys. {\bf10} (8) 361
\end{thebibliography}
\end{document}